\documentclass{article}

\usepackage{amsmath,amssymb,latexsym,graphicx} \usepackage[latin1]{inputenc}
\usepackage{times,fullpage}

\newcommand{\nc}{\newcommand}

\makeatletter \renewcommand{\@seccntformat}[1]{{\normalfont\bfseries{\csname
      the#1\endcsname}\hspace{0.5em}}} \renewcommand{\section}{\@startsection
        {section}%
        {1}%
        {0mm}%
        {-1.5\baselineskip}%
        {\baselineskip}%
        {\normalfont\normalsize\bfseries\centering}%
        } \renewcommand{\subsection}{\@startsection
        {subsection}%
        {2}%
        {0mm}%
        {-0.5\baselineskip}%
        {-0.5em}%
        {\normalfont\normalsize\bfseries}%
        } \renewcommand{\@makecaption}[2]{\begin{center}#1. #2\end{center}}
      \makeatother

\newtheorem{Theorem}{Theorem}
\newtheorem{Lemma}[Theorem]{Lemma} \newtheorem{Corollary}[Theorem]{Corollary}
\newtheorem{Proposition}[Theorem]{Proposition}
 \newtheorem{ex}[Theorem]{Example}
\newtheorem{de}[Theorem]{Definition} \newtheorem{re}[Theorem]{Remark}

\newenvironment{Example}{\begin{ex}\upshape}{\end{ex}}
\newenvironment{Definition}{\begin{de}\upshape}{\end{de}}
\newenvironment{Remark}{\begin{re}\upshape}{\end{re}}

\newcommand{\proof}{\medskip\noindent\textit{Proof: }}
\newcommand{\proofend}{\unskip\nobreak\hfill\raisebox{-.067em}{\large$\Box$}\vspace{\topsep}\par}
\nc{\fin}{ \proofend} \nc{\fine}{\eqno\Box}
\nc{\fina}{\eqno\makebox[0pt]{\hspace{-\platz}$\Box$}}

\newcommand{\IF}{\textbf{if}} \newcommand{\THEN}{\textbf{then}}
\newcommand{\ELSE}{\textbf{else}} 
 \newcommand{\FOR}{\textbf{for}}
\newcommand{\FORALL}{\textbf{for all}} \newcommand{\TO}{\textbf{to}}
\newcommand{\DO}{\textbf{do}} 
  
 \newcommand{\AND}{\textbf{and}}
\newcommand{\ACCEPT}{\underline{accept}}
\newcommand{\REJECT}{\underline{reject}} \newcounter{zeilencounter}
\newcommand{\zeile}{\stepcounter{zeilencounter}\>\hspace{4mm}{\small\itshape\arabic{zeilencounter}}\>}

\nc{\PP}[4]{\begin{center}\normalfont\fbox{\fbox{\textit{#1}}\hspace{5mm}\begin{tabular}[t]{rp{6.7cm}}\textit{Input:}&#2.\\\textit{Parameter:}&#3.\\\textit{Question:}&#4?\end{tabular}}\end{center}\noindent}

\nc{\PPP}[3]{\begin{center}\normalfont\fbox{\begin{tabular}[t]{rp{8cm}}\textit{Input:}&#1.\\\textit{Parameter:}&#2.\\\textit{Question:}&#3?\end{tabular}}\end{center}\noindent}

\nc{\PR}[3]{\begin{center}\normalfont\fbox{\fbox{\textit{#1}}\hspace{5mm}\begin{tabular}[t]{rp{7cm}}\textit{Input:}&#2.\\\textit{Question:}&#3?\end{tabular}}\end{center}}

\nc{\bigmid}{\;\big|\;}

\newcommand{\str}[1]{\mbox{${\cal #1}$}}

\renewcommand{\phi}{\varphi} \renewcommand{\max}{\textup{max}}

\newcommand{\NN}{\text{$\mathbb N$}} 

\nc{\STR}{{\textit{STR}}} \nc{\ORD}{{\textit{ORD}}} \nc{\STW}{{\textit{STW}}}
\nc{\GRA}{{\textit{GRAPH}}} \nc{\MC}{{\textit{MC}}} \nc{\VC}{{\textit{VC}}}
\nc{\SI}{{\textit{SI}}} \nc{\DS}{{\textit{DS}}} \nc{\CLI}{{\textit{CLIQUE}}}
\nc{\AM}{{\textit{AM}}} \nc{\NM}{{\textit{NM}}}

\nc{\FO}{\textup{FO}} \nc{\EFO}{\textup{EFO}} \nc{\AFO}{\textup{AFO}}
\nc{\MSO}{\textup{MSO}} \nc{\LFP}{\textup{LFP}} \nc{\FPT}{\textup{FPT}}
\nc{\PTIME}{\textup{PTIME}} \nc{\NP}{\textup{NP}} \nc{\W}{\textup{W}}
\nc{\A}{\textup{A}}

\nc{\Pow}{\textup{Pow}} \nc{\cost}{\textup{cost}} \nc{\opt}{\textup{opt}}
\nc{\MIN}{\textup{MIN}} \nc{\MAX}{\textup{MAX}} \nc{\PB}{\textup{PB}}
\nc{\Sat}{\textup{Sat}} \nc{\APX}{\textup{APX}}

\nc{\tw}{\textup{tw}} \nc{\ltw}{\textup{ltw}} \nc{\fp}{\textup{fp}}

\nc{\HOM}{{\textit{HOM}}} \nc{\EMB}{{\textit{EMB}}} \nc{\GW}{\textit{GW}}
\nc{\var}{{\textup{var}}}

\nc{\prm}{\le^{\text{\normalfont fp}}_{\text{\normalfont m}}}
\nc{\prT}{\le^{\text{\normalfont fp}}_{\text{\normalfont T}}}
\nc{\eprT}{\equiv^{\text{\normalfont fp}}_{\text{\normalfont T}}}
\nc{\pprm}{\le^{\text{\normalfont fpp}}_{\text{\normalfont m}}}
\nc{\eprm}{\equiv^{\text{\normalfont fp}}_{\text{\normalfont m}}}
\nc{\eptrm}{\equiv^{\text{\normalfont fpp}}_{\text{\normalfont m}}}
\nc{\mclass}[1]{[#1]^{\text{\normalfont fp}}_{\text{\normalfont m}}}

\newcounter{algcounter}
\newcommand{\alg}{\refstepcounter{algcounter}\begin{center}Algorithm
    \arabic{algcounter}\end{center}}

\begin{document}
\title{\large\bfseries Fixed-Parameter Tractability, Definability,  and Model Checking}
\author{\normalsize J\"org Flum
\thanks{Institut f\"ur Mathematische Logik, Eckerstr.\ 1,
79104 Freiburg, Germany. Email: \texttt{flum@sun2.ruf.uni-freiburg.de}}
\and\normalsize Martin Grohe
\thanks{Department of Mathematics, Statistics, and Computer
Science, University of Illinois at Chicago,
851 S. Morgan St.\ (M/C 249), Chicago, IL 60607-7045, USA. 
Email: \texttt{grohe@uic.edu}}} 
\date{\normalsize\today}
\maketitle

\begin{abstract}
  In this article, we study parameterized complexity theory from the
  perspective of logic, or more specifically, descriptive complexity theory.
  
  We propose to consider parameterized \emph{model-checking} problems for
  various fragments of first-order logic as generic parameterized problems and
  show how this approach can be useful in studying both fixed-parameter
  tractability and intractability. For example, we establish the equivalence
  between the model-checking for existential first-order logic, the
  homomorphism problem for relational structures, and the substructure
  isomorphism problem. Our main tractability result shows that model-checking
  for first-order formulas is fixed-parameter tractable when restricted to a
  class of input structures with an excluded minor. On the intractability
  side, for every $t\ge 0$ we prove an equivalence between model-checking for
  first-order formulas with $t$ quantifier alternations and the parameterized
  halting problem for alternating Turing machines with $t$ alternations. We
  discuss the close connection between this \emph{alternation hierarchy} and
  Downey and Fellows' W-hierarchy.
  
  On a more abstract level, we consider two forms of definability, called
  \emph{Fagin definability} and \emph{slicewise definability}, that are
  appropriate for describing parameterized problems. We give a
  characterization of the class FPT of all fixed-parameter tractable problems
  in terms of slicewise definability in finite variable least fixed-point
  logic, which is reminiscent of the Immerman-Vardi Theorem characterizing
  the class PTIME in terms of definability in least fixed-point logic.
\end{abstract}

\section{Introduction}\label{sec:1}
Parameterized complexity is a branch of complexity theory which has matured in
the last 10 years, as witnessed in the culminating monograph \cite{dowfel99}.
It gives a framework for a refined complexity analysis of hard algorithmic
problems.  The basic idea can best be explained by an example: Consider the
problem of evaluating a query in a relational database. This problem usually
has a high complexity (depending on the query language, of course, but the
problem is NP-complete even for the very basic conjunctive queries
\cite{chamer77}). The main factor contributing to this complexity is the
length of the query. In practice, however, queries are usually short,
certainly much shorter than the size of the database. Thus when analyzing the
complexity of the problem we should put much more emphasis on the size of the
database than on the length of the query. An algorithm evaluating a query of
length $k$ in a database of size $m$ in time $O(2^k\cdot m)$ is therefore much
better than one performing the same task in time $O(m^{k/2})$, although both
are exponential.

Parameterized complexity theory studies problems whose instances are
\emph{parameterized} by some function of the input, such as the length of the
query in our example. The idea is to choose the parameterization in such a way
that it can be assumed to take small values for the instances one is
interested in.  Then the complexity of an algorithm is measured not only in
the size of the input, but also in terms of the parameter. A parameterized
problem is \emph{fixed-parameter tractable} if there is an algorithm solving
it in time $f(k)\cdot n^c$, where $n$ denotes the size of the input, $k$ the
parameter, and $f:\mathbb N\rightarrow\mathbb N$ is a computable function and
$c>0$ a constant.

Parameterized complexity theory provides methods for proving problems to be
fixed-parameter tractable, but also gives a framework for dealing with
apparently intractable problems in a similar way that the theory
of NP-completeness does in classical complexity theory.

\medskip The purpose of this article is to establish a very fruitful connection
between parameterized complexity theory and logic. Our approach is
that of descriptive complexity theory. We study the definability of
parameterized problems and try to obtain information about the
parameterized complexity of the problems through the syntactical
structure of the defining sentences. On the one hand, we use this approach to
prove that certain problems are tractable because they can be defined
by syntactically simple formulas. On the other hand, we
characterize classes of intractable problems by syntactical means.

Central to our approach are \emph{parameterized model-checking}
problems of the following form. For a class $\Phi$ of formulas, we let
$\MC(\Phi)$ be the problem

\PP{$\MC(\Phi)$}{A finite structure $\mathcal A$}{A sentence
  $\phi\in\Phi$}{Does $\mathcal A$ satisfy $\phi$}

\noindent
In most cases, $\Phi$ will be a fragment of first-order logic. 

After a preliminary section, we discuss some basic facts about
parameterized model-checking problems in Section \ref{sec:3}. In Section
\ref{sec:4} we introduce two notions of definability of parameterized
problems, which we call slicewise definability and Fagin definability,
and relate them to model-checking. We then show how Fagin definability
can be used to establish the fixed-parameter tractability of various
problems. 

In Section \ref{sec:5} we study the parameterized complexity of the
model-checking problem for $\Sigma_1$-formulas (that is, existential
first-order formulas in prenex normal form). We associate a graph with each
such formula and use it to establish a surprisingly close connection between
this model-checking problem, the homomorphism problem, and the subgraph
isomorphism problem. As an application of our result we show that for
$\Sigma_1$-sentences whose graph has bounded tree-width the model-checking
problem is fixed-parameter tractable, even if inequalities are disregarded in
the graph of the formula. Model-checking for formulas with a tree-like graph
or hypergraph has recently received much attention (see
\cite{cheraj97,kolvar98,gotleosca99,flufrigro01}).

So far we have only looked for tractable cases of the model-checking problem
$\MC(\Phi)$ that are obtained by restricting the class of formulas $\Phi$. A
different approach is to restrict the class of structures where the input
structure $\mathcal A$ is taken from (see, for example,
\cite{cou90,see96,frigro99}). We prove a far reaching result: For any class
$C$ of graphs with an excluded minor, the model-checking problem for
first-order logic is fixed-parameter tractable if the inputs are taken from
$C$.  This implies, for example, that parameterized versions of the dominating
set problem or the (induced) subgraph isomorphism problem are fixed-parameter
tractable when restricted to such classes of graphs.

Our last result on fixed-parameter tractability is a descriptive
characterization of the complexity class FPT of all fixed-parameter tractable
problems in terms of slicewise definability in finite variable fragments of
least-fixed point logic.  This simple result can be seen as a parameterized
analogue of the well-known Immerman-Vardi Theorem \cite{imm86,var82}
characterizing the class PTIME in terms of definability in least-fixed-point
logic.

The final section is devoted to fixed-parameter intractability.  We
define a hierarchy $\A[t]$ of parameterized complexity classes in terms
of alternating Turing machine acceptance ($t$ is the number of alternations).
This hierarchy can be seen as a parameterized analogue of the polynomial
hierarchy.  We prove that for all $t\ge 1$, the model-checking problem for
$\Sigma_t$-formulas is complete for the $t$th level of this hierarchy.
Then we study the relation between our $\A$-hierarchy and Downey and
Fellows' $\W$-hierarchy. It is known that the first levels of the
respective hierarchies, $\A[1]$ and $\W[1]$ coincide
\cite{caichedowfel97}. We slightly improve a result of Downey, Fellows,
and Regan \cite{dowfelreg98} relating $\W[t]$, the $t$th level of
the \W-hierarchy, to the model-checking problems for a certain
fragment of $\Sigma_t$. However, the questions whether $\text{A}[t]$
and $\text{W}[t]$ coincide for $t\ge 2$ remains open.

\section{Preliminaries}\label{sec:2}

\subsection{Logic}
We assume that the reader is familiar with first-order logic; we just recall a
few basic notions to fix our notation (compare \cite{ebbflu95} for a more detailed introduction of these notions).

\emph{In this article, a \emph{vocabulary} is a finite set of relation
  symbols.} Associated with every relation symbol is a natural number,
its \emph{arity}. The \emph{arity} of a vocabulary is the maximal
arity of the relation symbols it contains. Usually, vocabularies are
also permitted to contain function and constant symbols. All results
of this article, with the single exception of Theorem \ref{theo:15}, would
remain true if function and constant symbols were allowed, but adding
them would not give us any new insights. So, for convenience,  we
restrict our attention to relational vocabularies.  In the following,
$\tau$ always denotes a vocabulary.

A {\em $\tau$-structure} $\mathcal A$ consists of a set $A$, called the {\em
  universe} of $\mathcal A$, and a relation $R^{\mathcal A}\subseteq A^r$ for
each $r$-ary relation symbol $R\in\tau$. We synonymously write $\bar
  a\in R^{\mathcal A}$ or $R^{\mathcal A}\bar a$ to denote that the
  tuple $\bar a\in A^r$ belongs to the relation $R^{\mathcal A}$. For
$\tau\subseteq\tau'$, a $\tau$-structure $\mathcal A$ is the
\emph{$\tau$-reduct} of a $\tau'$-structure $\mathcal A'$ if $A=A'$ and
$R^{\mathcal A}=R^{\mathcal A'}$ for all $R\in\tau$. A
$\tau'$-structure $\mathcal A'$ is a \emph{$\tau'$-expansion} of a
$\tau$-structure $\mathcal A$ if $\mathcal A$ is the \emph{$\tau$-reduct} of
$\mathcal A'$. 

\emph{We only consider finite structures.} When we consider
classes of structures, they are always assumed to be closed under isomorphism.
$\STR$ denotes the class of all (finite) structures. If $C$ is a class of
structures, $C[\tau]$ denotes the subclass of all $\tau$-structures in
$C$. Furthermore, $C[s]$ denotes the class of all structures in $C$ whose
vocabulary is at most $s$-ary. We
consider graphs as $\{E\}$-structures $\mathcal G=(G,E^{\mathcal G})$, where
$E^{\mathcal G}$ is an irreflexive and symmetric binary relation (i.e.\ 
graphs are loop-free and undirected). \GRA\ denotes the class of all graphs.

The class of all first-order formulas is denoted by $\FO$. Recall that
\emph{atomic formulas} are formulas of the form $x=y$ or $Rx_1\ldots x_r$,
where $x,y,x_1,\ldots,x_r$ are variables and $R$ is a $r$-ary relation symbol.
\emph{Literals} are atomic or negated atomic formulas. A first-order
formula $\phi$ is in \emph{negation normal form} if negation symbols only
occur directly in front of atomic subformulas. $\phi$ is
\emph{existential} (\emph{universal}) if it is in negation normal form
and contains no universal quantifiers (no existential quantifiers,
respectively). $\phi$ is in \emph{prenex normal form} if it is of the
form $Q_1x_1\ldots Q_kx_k\theta$, where
$Q_1,\ldots,Q_k\in\{\exists,\forall\}$ and $\theta$ is quantifier-free. 

\EFO\ (\AFO) denotes the class of all existential (universal,
respectively) first-order formulas.  For $t\ge 1$, $\Sigma_t$ denotes
the class of all \FO-formulas of the form
\[
\exists x_{11}\ldots\exists x_{1k_1}\forall x_{21}\ldots\forall
x_{2k_2}\;\ldots\; Qx_{t1}\ldots Qx_{tk_t}\;\theta,
\]
where $Q=\forall$ if $t$ is even and $Q=\exists$ otherwise and $\theta$ is
quantifier-free. $\Pi_t$-formulas are defined analogously starting with
a block of universal quantifiers. 

If $\Phi$ is a class of formulas of some logic, then $\Phi[\tau]$ denotes the
class of all formulas of vocabulary $\tau$ in $\text{L}$, and $\Phi[s]$ denotes
the class of all formulas in $\Phi$ whose vocabulary is at most $s$-ary. We
write $\mathcal A\models \varphi$ if, for some $\tau$, $\mathcal A$ is a
$\tau$-structure, $\varphi$ is in $\text{L}[\tau]$, and $\mathcal A$ is a
model of $\varphi$.

\subsection{Coding issues}
We use random access machines (RAMs) with the uniform cost measure as our
underlying model of computation (cf.\ \cite{ahohopull74}).

Very often, the objects of our computations are structures. Therefore, we have
to fix a way of representing structures on a RAM. The two most common ways of
doing this are the \emph{array representation} and the \emph{list
  representation}. For both representations we assume that the universes of our
structures are  initial segments of the natural numbers; of course this is no
real restriction because every structure is isomorphic to one with such a
universe. 

Both representations start with an encoding of the vocabulary and a
natural number representing the size of the universe of the structure.
The difference between the two representations is in how relations are
stored. In the array representation, a $k$-ary relation is stored as a
$k$-dimensional array with $0,1$-entries. For graphs, this is just the
adjacency matrix. The advantage of this representation is that for
each tuple it can be checked in constant time whether it belongs to
the relation or not. However, for sparse relations this representation
wastes a lot of space.

In the more concise list representation, a relation is represented as a list
of all tuples it contains. Clearly, the list representation of a structure can
be computed from the array representation in linear time, but not vice versa.
For graphs $\mathcal G$, it is easy to construct the common adjacency
list representation from the
list representation (in time linear in $|G|+(\text{size of the
  representation})$, where $|G|$ denotes the number of elements in $G$).  In this article, we always assume that structures are
given in the list representation, but all results also hold for the array
representation.  The \emph{size} of a structure $\mathcal A$, denoted by
$||\mathcal A||$, is defined to be $|A|+(\text{size of the list representation
  of }\mathcal A)$. The complexity of algorithms on
structures is measured in this size. Remark \ref{l4} shows that this can be relevant.

\subsection{Parameterized problems}
We only recall those notions of the theory needed in this article. For a
comprehensive treatment we refer the reader to Downey and Fellow's
recent monograph \cite{dowfel99}.  A \emph{parameterized problem} is a
set $P\subseteq\Sigma^*\times\Pi^*$, where $\Sigma$ and $\Pi$ are
finite alphabets. Following \cite{dowfel99}, we usually represent a
parameterized problem $P$ in the following form:

\PP{P}{$x\in\Sigma^*$}{$y\in\Pi^*$}{Is $(x,y)\in P$}

\noindent
In most cases, we have $\Pi=\{0,1\}$ and consider the parameters
$y\in\Pi^*$ as natural numbers (in binary).
A natural example is the parameterized version of the well-known
\textit{VERTEX COVER} problem:

\PP{VC}{Graph $\mathcal G$}{$k\in\mathbb N$}{Does $\mathcal G$ have a vertex
  cover of size $k$}

\noindent
Recall that a \emph{vertex cover} of a graph is a set $X$ of vertices such
that every edge is incident to one of the vertices in $X$. Similarly, we can
define parameterized versions of \emph{DOMINATING SET (DS)} and
\textit{CLIQUE}. (A \emph{dominating set} of a graph is a set $X$ of vertices
such that every vertex not contained in $X$ is adjacent to a vertex in $X$. A
\emph{clique} is a set of pairwise adjacent vertices.)

An example where the set of parameters is not $\mathbb N$, but the
class \GRA\ of all finite graphs is the following parameterized
\textit{SUBGRAPH ISOMORPHISM} problem:

\PP{SI}{Graph $\mathcal G$}{Graph $\mathcal H$}{Is $\mathcal H$
  isomorphic to a subgraph of $\mathcal G$}
  
\noindent
Similarly, we can define  parameterized versions of the
\textit{INDUCED SUB\-GRAPH ISOMORPHISM} problem and the \textit{GRAPH
  HOMOMORPHISM} problem.

\begin{Definition}\label{def:1}
  A parameterized problem $P\subseteq\Sigma^*\times\Pi^*$ is
  \emph{fixed-parameter trac\-ta\-ble} if there is a computable function
  $f:\Pi^*\rightarrow\mathbb N$, a constant $c\in\mathbb N$, and an algorithm
  that, given a pair $(x,y)\in\Sigma^*\times\Pi^*$, decides if $(x,y)\in
  P$ in time $f(|y|)\cdot |x|^c$.\footnote{This is what Downey and
  Fellows call \emph{strongly uniformly fixed-parameter tractable}. 
  For variants of this definition, and also of Definition \ref{def:2}
  and the definition of the \W-hierarchy, the reader should consult
  \cite{dowfel99}.}
  
We denote the class of all fixed-parameter tractable problems by
  \FPT.
\end{Definition}

Of course we can always consider parameterized problems as classical problems
and determine their complexity in the classical sense. Clearly, every
parameterized problem in PTIME is also in FPT.

The best currently known algorithm for vertex cover \textit{VC} has running time
$O(k\cdot n+\max\{1.255^k\cdot k^2,1.291^k\cdot k\})$ \cite{felste99}, where $n$ denotes the size
of the input graph. Thus $\textit{VC}\in\FPT$. 

\subsection{Reductions between parameterized problems}
It is conjectured that none of the problems \textit{DS}, \textit{CLIQUE},
\textit{SI} is in \FPT.  As it is often the case in complexity theory, we can
not actually prove this, but only prove that the problems are hard for certain
complexity classes that are conjectured to contain {\FPT} strictly.  To do
this we need a suitable concept of reduction. We actually introduce
three different types of reduction:

\begin{Definition}\label{def:2}
  Let $P\subseteq\Sigma^*\times\Pi^*$ and $P'\subseteq
  (\Sigma')^*\times(\Pi')^*$ be parameterized problems.
\begin{enumerate}
\item A \emph{parameterized T-reduction} from $P$ to $P'$ is an
  algorithm with an oracle for $P'$ that solves any instance $(x,y)$ of
  $P$ in time $f(|y|)\cdot |x|^c$ in such a way that for  all  questions
  $(x',y')\in P'?$ to the oracle we have $|y'|\le g(|y|)$ (for
  computable functions $f,g:\mathbb N\rightarrow\mathbb N$ and a
  constant $c\in\mathbb N$).

$P$ is {\em fixed-parameter T-reducible} to $P'$ (we write $P\prT P'$),
if there is a parameterized T-reduction from $P$ to
$P'$.
\item A \emph{parameterized m-reduction} from $P$ to $P'$ is an
  algorithm that computes for every instance $(x,y)$ of
  $P$ an instance $(x',y')$ of $P'$ in time $f(|y|)\cdot |x|^c$ such that
  $|y'|\le  g(|y|)$  and
\[
(x,y)\in P\iff(x',y')\in P'
\]
(for computable functions $f,g:\mathbb N\rightarrow\mathbb N$ and a
constant $c\in\mathbb N$).

$P$ is {\em fixed-parameter m-reducible} to $P'$ (we write $P\prm P'$),
if there is a parameterized m-reduction from $P$ to
$P'$.
\end{enumerate}
\end{Definition}

Whereas every parameterized problem that is in \PTIME\ (when considered
as a classical problem) is in \FPT, it is not the case that every
\PTIME\ many-one reduction between two parameterized problems is also
a parameterized m-reduction. To capture both concepts we occasionally
use the following third kind of reduction:

\begin{Definition}\label{def:2a}
  Let $P\subseteq\Sigma^*\times\Pi^*$ and $P'\subseteq
  (\Sigma')^*\times(\Pi')^*$ be parameterized problems.

A \emph{pp m-reduction} from $P$ to $P'$ is a parameterized
m-reduction from $P$ to $P'$ that is also a polynomial time many-one reduction from $P$ to $P'$ in the classical
sense, i.e.\ the function $f$ in Definition \ref{def:2}(2) is a polynomial.

$P$ is {\em pp m-reducible} to $P'$ (we write $P\pprm P'$),
if there is a pp m-reduction from $P$ to
$P'$.
\end{Definition}

For example, $\CLI\prm\SI$\/ by the simple parameterized
m-reduction that reduces the instance $(\mathcal G,k)$ of \CLI\ to the
instance $(\mathcal G,\mathcal K_k)$ of $\SI$. Here $\mathcal K_k$
denotes the complete graph with $k$ vertices. Note that if we represent
integers in binary, this reduction is not a pp m-reduction.

Observe that $\prT$, $\prm$, and $\pprm$ are transitive and that for all $P,P'$
we have
\[
P\pprm P'\implies P\prm P'\quad\text{and}\quad P\prm P'\implies P\prT P'.
\]
Furthermore, if $P\prT P'$ and $P'\in\FPT$ then
$P\in\FPT$. For any of the reductions $\prT,\prm,\pprm$ we let
$\equiv^{\cdots}_{\ldots}$ denote the corresponding equivalence relation.

We define \emph{hardness} and \emph{completeness} of parameterized
problems for a parameterized complexity class (under parameterized
m- or T-reductions) in the usual way.
For a parameterized problem $P$, we let $\mclass{P}:=\{P'\mid P'\prm
P\}$, and for a class P of parameterized problems $\mclass{\text{P}}:=\bigcup_{P\in\text{P}}\mclass{P}$. 

\begin{Remark}\label{l1}
  Very often, it is natural to think of a parameterized problem $P$ as
  derived from a (classical) problem $L\subseteq\Sigma^*$ by a
  \emph{parameterization} $p:\Sigma^*\rightarrow\mathbb N$ in such a
  way that $P=\{(x,k)\mid x\in L,k=p(x)\}$. 
  
  Slightly abusing notation, we represent such a $P$ in the form
  
  \PP{P}{$x\in\Sigma^*$}{$p(x)$}{Decide if $x\in L$}

  \noindent
  As an example, let us reconsider the subgraph isomorphism
  problem. Instead of taking graph $\mathcal H$ as the parameter, we
  may also consider pairs of graphs $(\mathcal G,\mathcal H)$ as inputs
  and parameterize the problem by the size of $\mathcal H$. In our new
  notation, this would be the problem

\PP{${\SI}\,'$}{Graphs $\mathcal G$, $\mathcal H$}{$||\mathcal H||$}{Is 
$\mathcal H$
isomorphic to a subgraph of $\mathcal G$}

\noindent
It is easy to see, however, that $\SI\eprm\SI\,'$.
\end{Remark}

\subsection{Parameterized intractability}
Some combinatorial problems are
provably not fixed-parameter tractable, and others, such as \textit{GRAPH
  COLORABILITY}, are not fixed-parameter tractable unless $\PTIME=\NP$.
However, many interesting problems, such as the parameterized \textit{CLIQUE}
problem, do not seem to be fixed-parameter tractable, although there is no
known way to prove this or reduce it to classical complexity theoretic
questions such as $\PTIME\stackrel{?}{=}\NP$.
To classify such problems, Downey and Fellows (cf.\
\cite{dowfel99}) introduced a hierarchy
\[
\textup W[1]\subseteq \textup W[2]\subseteq\cdots
\]
of classes above $\FPT$. These classes can best be defined in terms of the
satisfiability problem for formulas of propositional logic. Formulas of
propositional logic are build up from \emph{propositional variables}
$X_1,X_2,\ldots$ by taking conjunctions, disjunctions, and negations. The
negation of a formula $\phi$ is denoted by $\neg\phi$. We distinguish between
\emph{small conjunctions}, denoted by $\wedge$, which are just conjunctions of
two formulas, and \emph{big conjunctions}, denoted by $\bigwedge$, which are
conjunctions over arbitrary finite sets of formulas. Analogously, we
distinguish between \emph{small disjunctions}, denoted by $\vee$, and
\emph{big disjunctions}, denoted by $\bigvee$.

Every formula $\phi$ corresponds to a labeled tree $\mathcal T_\phi$ in a
natural way. The \emph{size} of $\phi$ is defined to be the number of vertices
of $\mathcal T_\phi$. The \emph{depth} of $\phi$ is defined to be the maximum
number of nodes labeled $\wedge,\bigwedge,\vee,\bigvee$ on a path from the
root to a leaf of $\mathcal T_\phi$. Thus when computing the depth, we do not
count negations.

A formula is \emph{small} if it only contains small conjunctions and small
disjunction. We define $C_0=D_0$ to be the class of all small formulas. For an
$i\ge 1$, we define $C_{i}$ to be the class of all big conjunctions of
formulas in $D_{i-1}$, and we define $D_{i}$ to be the class of all big
disjunctions of formulas in $C_{i-1}$. Note that these definitions are purely
syntactical; every formula formula in a $C_i$ or $D_i$ is equivalent to a
formula in $C_0$. But of course the translation from a formula in $C_i$ to an
equivalent formula in $C_0$ usually increases the depth of a formula. For all
$i,d\ge0$ we let $C_{i,d}$ denote the class of all formulas in $C_i$ whose
small subformulas have depth at most $d$ (equivalently, we may say that the
whole formula has depth at most $d+i$). We define $D_{i,d}$ analogously.

The \emph{weight} of an assignment $\alpha$ for the variables of a
propositional formula is the number of variables set to \textsc{True} by
$\alpha$. For any class $P$ of propositional formulas, let \emph{weighted
  satisfiability for $P$} be the following parameterized problem:

\PP{$\text{WSAT}(P)$}{$\phi\in P$}{$k\in\mathbb N$}{Does $\phi$ have a
  satisfying assignment of weight $k$}

Now we are ready to define the \emph{W-hierarchy}: For every $t\ge 1$, we let
\[
\W[t]:=\bigcup_{d\ge0}\left[\textit{WSAT}(C_{t,d})\right]^{\fp}_{\text{m}}.
\]
In other words, a parameterized problem is in $\W[t]$ if there is a
$d\ge 0$ such that the problem is fixed-parameter m-reducible to the weighted
satisfiability problem for $C_{t,d}$. 
It is an immediate consequence of the definition of parameterized m-reductions
that $\FPT\subseteq\W[1]$. Actually, it is conjectured that this inclusion is
strict and that $\W[t]$ is strictly contained in $\W[t+1]$ for every $t\ge 1$.

\begin{Example}
  The parameterized \CLI-problem is in $\W[1]$. To see this, for every graph
  $\mathcal G$ we describe a propositional formula $\phi:=\phi({\mathcal
    G})\in C_{1,1}$ such that $\mathcal G$ has a clique of size $k$ if, and
  only if, $\phi$ has a satisfying assignment of weight $k$. It
  will be obvious from the construction that $\phi$ can be computed from
  $\mathcal G$ in polynomial time.

So let $\mathcal G$ be a graph. For all $a\in G$ let
$X_{a}$ be a propositional variable. Let
\[
\phi:=\bigwedge_{\substack{a,b\in G, a\neq b\\ab\not\in E^G}}(\neg X_{a}\vee
\neg X_b)
\]
Then every satisfying assignment of $\phi$ corresponds to a clique of $\mathcal G$.
\end{Example} 

Actually, Downey and Fellows proved the following non-trivial result:

\begin{Theorem}[Downey and Fellows \cite{dowfel95b,dowfel95}]
\begin{enumerate}
\item
\CLI\ is $\W[1]$-complete under parameterized m-reductions.
\item
\DS\ is $\W[2]$-complete under parameterized m-reductions.
\end{enumerate}
\end{Theorem}

\begin{Remark}
  Downey and Fellows phrase their definition of the W-hierarchy in terms of
  Boolean circuits rather than propositional formulas. But since the classes
  of the hierarchy only involve circuits/formulas of bounded depth, this does
  not really make a difference (cf.\ \cite{dowfel99}). In their definition of
  $\W[t]$, Downey and Fellows admit more complicated formulas than those in
  $C_t$. But they prove that our definition is equivalent. A surprising
  by-product of their results is that for every $t\ge 1$ and every $d\ge 0$,
  the problem $\textit{WSAT}[D_{t+1,d}]$ is contained in $\W[t]$. It is not
  hard to prove this result directly, and even easier to prove that
  $\textit{WSAT}[D_{1,d}]$ is in \FPT. (This explains why we only defined a
  hierarchy using the $C_t$s).
  
  There is another, more serious source of confusion in the various
  definitions of the W-hierarchy: Downey and Fellows are never really clear
  about what kind of reductions they are using to define the classes. We
  decided, more or less in accordance with \cite{dowfel99}, that parameterized
  m-reductions are most natural.
\end{Remark}

We will further discuss the $\W$-hierarchy and other seemingly
intractable classes in Section \ref{sec:9}.

\section{Model-checking}\label{sec:3}
In this article we are mainly concerned with the complexity of
various \emph{parameterized model-checking problems}. For a set $\Phi$ of
formulas, we let 
\[
\MC(\Phi):=\big\{(\mathcal A,\phi)\bigmid\mathcal
A\in\STR,\phi\text{ sentence in }\Phi,\mathcal A\models\phi\big\},
\]
or more intuitively, 

\PP{$\MC(\Phi)$}{$\mathcal A\in\STR$}{$\phi\in\Phi$}{Does $\mathcal
  A\models\phi$}

\noindent
In this section we collect a few basic facts about parameterized
model-checking problems. For every $\Phi$ we consider, we assume that we have
fixed an encoding $\gamma:\Phi\rightarrow\{0,1\}^*$, and we let $||\phi||$ be
the length of $\gamma(\phi)$.

Taking sentences as parameters seems a little
unusual. The following parameterization of the model checking problem looks
more natural:

\PP{$\MC\,'(\Phi)$}{$\mathcal A\in\STR$, $\phi\in\Phi$}{$||\phi||$}{Does
  $\mathcal A\models\phi$}

\noindent
However, it is easy to see that $\MC(\Phi)\eprm\MC\,'(\Phi)$.

It is well-known that various problems of model theory or complexity theory can be reduced from structures to graphs.
The following two lemmas contain such reductions. Although their proofs
only use standard techniques, they are
subtle and require some care. Therefore we decided
to give the proofs in some detail. We will apply these lemmas several
times later.

\begin{Lemma}\label{lem:1}
  There are polynomial time transformations
  that associate with every structure $\mathcal A\in\STR$ a graph
  $\mathcal H(\mathcal A)$ and with every sentence $\phi\in\FO$ a sentence
  $\phi_{\GRA}\in\FO$, respectively, such that
\[
\mathcal A\models\phi\iff\mathcal H(\mathcal A)\models\phi_{\GRA}.
\]
Furthermore for every $t\ge 1$, if $\phi\in\Sigma_t$ then
$\phi_{\GRA}\in\Sigma_{t+1}$ and if $\phi\in\Pi_t$ then
$\phi_{\GRA}\in\Pi_{t+1}$.
\end{Lemma}

\proof
Let $\mathcal A\in\STR[\tau]$ and
$\phi\in\FO[\tau]$. Without loss of generality we can assume that
$\phi$ is in prenex and in negation normal form. 

\medskip
\textit{Step 1.}
In the first step we translate $\mathcal A$ to a structure
$\mathcal B(\mathcal A)$ of a vocabulary $\beta(\tau)$ that only
consists of unary and binary relation symbols. $\phi$ is translated to
a corresponding sentence $\phi_B$ of vocabulary $\beta(\tau)$.

$\beta(\tau)$ contains unary relation symbols $U$ and $U_R$ for each
symbol $R\in\tau$ and binary relation
symbols $E_1,\ldots,E_s$, where $s$ is the arity of $\tau$.

The universe of $\mathcal B(\mathcal A)$ is
\[
B(\mathcal A):=A\cup\{b(R,\bar a)\mid R\in\tau,\bar a\in R^{\mathcal
  A}\}.
\]
We assume that the elements $b(R,\bar a)$ are all pairwise
distinct and distinct from those in $ A$.
Note that the cardinality of $B(\mathcal A)$ is essentially $||\mathcal A||$,
up to an additive term depending on $\tau$. The unary relations
are defined in the obvious way: We let $U^{\mathcal B(\mathcal A)}:=A$
and $U_R^{\mathcal B(\mathcal A)}:=\{b(R,\bar a)\mid\bar a\in
R^{\mathcal A}\}$ for every $R\in\tau$. The binary relations
$E_1,\ldots,E_s$ are defined by
\[
E_i^{\mathcal B(\mathcal A)}:=\{(a_i,b(R,\bar a)),(b(R,\bar a),a_i)\mid R\in\tau,\bar
a=(a_1,\ldots,a_r)\in R^{\mathcal A}, 1\le i\le r\}.
\]
Note that $E_i^{\mathcal B(\mathcal A)}$ is symmetric,
this will be useful later.

To define $\phi_B$, we first relativize all quantifiers to $U$, i.e.\ 
we inductively replace all subformulas $\exists x\psi$ by $\exists
x(Ux\wedge\psi)$ and all subformulas $\forall x\psi$ by $\forall
x(Ux\rightarrow\psi)$. We obtain a formula $\phi'$.

$\phi_B$ is obtained from $\phi'$ by replacing every atomic subformula
$R\bar x$, for $r$-ary $R\in\tau$, by 
\begin{equation}\label{eq:voc1}
\exists
z(U_Rz\wedge\bigwedge_{i=1}^rE_ix_iz)
\end{equation}
where $z$ is a new variable. 
Then we have
\begin{equation}\label{eq:3}
\mathcal A\models\phi\iff\mathcal B(\mathcal A)\models\phi_B.
\end{equation}
Furthermore, $\mathcal B(\mathcal A)$ can be computed from $\mathcal
A$ in time $O(||\tau||\cdot||\mathcal A||)$, where $||\tau||$ denotes
the length of the encoding of $\tau$, and $\phi_B$ can be computed
from $\phi$ in linear time.

\medskip 
\textit{Step 2.}  In this step we replace the binary
relations $E_1,\ldots,E_s$ by a single new binary relation $E$. We let
\[
\gamma(\tau):=(\beta(\tau)\setminus\{E_1,\ldots,E_s\})\cup\{E,P_1,\ldots,P_s\},
\]
where $P_1,\ldots,P_s$ are new unary relation symbols.

We transform $\mathcal B(\mathcal A)$ to a $\gamma(\tau)$-structure
$\mathcal C(\mathcal A)$ as follows: For $1\le i\le s$ and for all
$a,b\in B(\mathcal A)$ such that $E_i^{\mathcal B(\mathcal A)}ab$ we
introduce a new element $c(i,a,b)$, add the pairs
$(a,c(i,a,b))$, $(c(i,a,b),a)$, $(c(i,a,b),b)$, $(b,c(i,a,b))$ to
$E^{\mathcal C(\mathcal A)}$ and add $c(i,a,b)$ to $P_i^{\mathcal C(\mathcal
  A)}$.

We define a sentence $\phi_C$ by replacing each subformula of $\phi_B$
of the form $E_ixy$ by $\exists z(Exz\wedge Ezy\wedge P_iz)$.
Then we have the analogue of \eqref{eq:3} and the subsequent
remarks for $\mathcal C(\mathcal A),\phi_C$ instead of $\mathcal
B(\mathcal A),\phi_B$.

\medskip \textit{Step 3}.  The restriction of $\mathcal C(\mathcal A)$
to $E$ is already a graph, i.e.\ $E^{\mathcal C(\mathcal A)}$ is
symmetric and irreflexive, so all we have to do is to eliminate the
unary relations $U,(U_R)_{R\in\tau},(P_i)_{1\le i\le s}$. Say,
$Q_1,\ldots,Q_l$ is an enumeration of all these relations.  Note that
every $a\in C(\mathcal A)$ is either isolated or of valence two
or adjacent to a vertex of valence two. We use this to define certain trees
$\mathcal T_1,\ldots,\mathcal T_l$ and corresponding existential
first-order formulas $\xi_1(x),\ldots,\xi_l(x)$ and attach a copy of
$\mathcal T_i$ to each vertex in $Q_i^{\mathcal C(\mathcal A)}$ in
such a way that in the resulting graph $\mathcal H(\mathcal A)$ we
have for all vertices $a$:
\[
\mathcal H(\mathcal
A)\models\xi_i(a)\iff a\in Q_i^{\mathcal C(\mathcal A)}.
\]
Furthermore, the $\mathcal T_i$ and thus the $\xi_i$ can be chosen of size
polynomial in $l$. We omit the details.

Then we let $\phi_{\GRA}$ be the formula obtained from $\phi_C$ by replacing
every subformula of the form $Q_ix$ by $\xi_i(x)$, for $1\le i\le l$, and
transforming the resulting formula into prenex normal form in the usual
manner. 

\medskip
Clearly the transformations $\mathcal A\mapsto\mathcal H(\mathcal A)$ and
$\phi\mapsto\phi_{\GRA}$ are polynomial, and we have
\[
\mathcal A\models\phi\iff\mathcal H(\mathcal A)\models\phi_{\GRA}.
\]
It remains to prove that if $\phi\in\Sigma_t$ ($\phi\in\Pi_t$) then
$\phi_{\GRA}\in\Sigma_{t+1}$ ($\phi_{\GRA}\in\Pi_{t+1}$, respectively). This
follows easily from the way we defined the sentences $\phi_B$,
$\phi_C$, and $\phi_{\GRA}$ in Steps 1--3, noting that
each positive (negative) occurrence of a relation symbol $R\in\tau$ only gives
rise to positive (negative, respectively) occurrences of $U_R$, the $E_i$,
 and the $P_i$.
Thus for the positive occurrences of the $R\in\tau$ we get a new block of
existential quantifiers and for the negative occurrences a new block of
universal quantifiers. This increases the alternation depth by at most one.
\proofend

Recall that for a class $C$ of structures and an $s\ge 0$, by  $C[s]$ we
denote the class of all structures in $C$ whose vocabulary is at most
$s$-ary. Similarly, for a class $\Phi$ of formulas, by $\Phi[s]$ we
denote the class of all formulas in $\Phi$ whose vocabulary is at most
$s$-ary.

\begin{Lemma}\label{lem:2}
  Let $s\ge 1$. There is a polynomial time transformation that
  associates with every structure $\mathcal A\in\STR[s]$ a graph
  $\mathcal H'(\mathcal A)$ and a linear time function that associates
  with every sentence $\phi\in\FO$ a sentence $\phi_{\GRA}'\in\FO$,
  such that
\[
\mathcal A\models\phi\iff\mathcal H'(\mathcal A)\models\phi_{\GRA}'.
\]
Furthermore, for all $t\ge 1$, if $\phi\in\Sigma_t$ then
$\phi_{\GRA}'\in\Sigma_{t}$ and if $\phi\in\Pi_t$ then
$\phi_{\GRA}'\in\Pi_{t}$.
\end{Lemma}

\proof 
Let us first look back at the proof of the last lemma and note that if
all relation symbols only occur positively in $\phi$ then the transformation
$\phi\mapsto\phi_{\GRA}$ only generates a new block of existential quantifiers.
If all relation symbols only occur negatively then we only get a new block of
universal quantifiers.  Thus if $\phi\in\Sigma_{2t-1}$ ($\phi\in\Pi_{2t}$) and
 all relation symbols only occur positively in $\phi$ then also
$\phi_{\GRA}\in\Sigma_{2t-1}$ ($\phi_{\GRA}\in\Pi_{2t}$, respectively).
Similarly, if $\phi\in\Sigma_{2t}$ ($\phi\in\Pi_{2t-1}$)  and
all relation symbols only occur negatively in $\phi$ then also
$\phi_{\GRA}\in\Sigma_{2t}$ ($\phi_{\GRA}\in\Pi_{2t-1}$, respectively).

We first transform $\mathcal A$ to an $\mathcal A'$
and $\phi$ to a $\phi'$ in the same prefix class such that 
\[
\mathcal A\models\phi\iff\mathcal A'\models\phi',
\]
and either all relation symbols in $\phi'$ occur positively or negatively,
whichever we need to apply the previous remark. For this purpose, let $\tau$ be an at most $s$-ary vocabulary. We let $\tau':=\tau\cup\{\overline{R}\mid R\in\tau\}$, where
$\overline{R}$ is a new relation symbol that has the same arity as
$R$. For every $\mathcal A\in\STR[\tau]$, we let ${\mathcal
  A}'$ be the $\tau'$-expansion of $\mathcal A$ with
$\overline{R}^{\mathcal A'}=A^r\setminus{R}^{\mathcal A}$ for 
$r$-ary $R\in\tau$. Note that $\mathcal A'$ can be computed from
$\mathcal A$ in time $O(||\mathcal A||^s)$.
To define $\phi'$ we either replace each negative literal $\neg
R\bar x$ by $\overline{R}\bar x$ or each positive literal $R\bar x$ by
$\neg\overline R\bar x$. 
\proofend

\begin{Remark}\label{l4}
  Note that if we represent structures by the array
  representation, then the transformation of Lemma \ref{lem:2} is actually
  polynomial even if we do not fix the arity of the vocabulary in advance.
  This follows from the fact that the array representation of the structure
  $\mathcal A'$ (in the proof of the lemma) can be computed from the array
  representation of $\mathcal A$ in linear time, uniformly over all
  vocabularies.
\end{Remark}

For a parameterized problem $P\subseteq\STR\times\Pi^*$ and a class
$C$ of structures we let $P|_C$ denote the \emph{restriction} of $P$
to $C$. In particular,
\[
\MC(\Phi)|_{\GRA}=\{(\mathcal G,\phi)\mid\mathcal
G\in\GRA,\phi\in\Phi,\mathcal G\models\phi\}.
\]
Note that for every class $\Phi$ of formulas and vocabulary $\tau$ the
two problems $\MC(\Phi[\tau])$ and $\MC(\Phi)|_{\STR[\tau]}$, though
formally different, are essentially the same. Therefore we do not
distinguish between them. 

\begin{Corollary}\label{cor:1}
\begin{enumerate}
\item $\MC(\FO)\eptrm\MC(\FO)|_{\GRA}$.
\item For all $t\ge 1$ we have
\[
\MC(\Sigma_{t})\pprm\MC(\Sigma_{t+1})|_{\GRA}
\quad\text{and}\quad
\MC(\Pi_{t})\pprm\MC(\Pi_{t+1})|_{\GRA}.
\]
\item For all $s\ge 2$, $t\ge 1$ we have
\[
\MC(\Sigma_{t}[s])\eptrm\MC(\Sigma_{t})|_{\GRA}
\quad\text{and}\quad
\MC(\Pi_{t}[s])\eptrm\MC(\Pi_{t})|_{\GRA}.
\]
\end{enumerate}
\end{Corollary}

We do not know whether $\MC(\Sigma_{t})\prm\MC(\Sigma_{t})|_{\GRA}$ and
$\MC(\Pi_{t})\prm\MC(\Pi_{t})|_{\GRA}$ for  $t\ge 1$.

\section{Defining parameterized problems}\label{sec:4}
Definability is the connection between arbitrary parameterized
problems and our logical analysis that focuses on model-checking problems.
In \cite{dowfelreg98}, Downey, Fellows, and Regan consider two forms of definability: 
Their exposition motivates two general notions of definability, which we call 
\emph{slicewise definability}
and \emph{Fagin definability}.

For a parameterized problem $P\subseteq\Sigma^*\times\Pi^*$ and $y\in\Pi^*$, we call $P\cap(\Sigma^*\times\{y\})$ the $y$th  \emph{slice} of $P$.

\begin{Definition}\label{def:3}
  Let $P\subseteq\STR\times\Pi^*$ be a parameterized problem and $\Phi$ a
  class of formulas. $P$ is \emph{slicewise $\Phi$-definable} if there is 
a
  computable function $\delta:\Pi^*\rightarrow\Phi$ such that for all
  $\mathcal A\in\STR$ and $y\in\Pi^*$ we have $(\mathcal A,y)\in P\iff\mathcal
  A\models\delta(y)$.
\end{Definition}

For example, the parameterized subgraph isomorphism problem $\SI$ is
slicewise $\Sigma_1$-definable via the function
$\delta:\GRA\rightarrow\Sigma_1$ defined as follows: For a graph $\mathcal 
H$
with vertex set $H=\{h_1,\ldots,h_n\}$ of cardinality $n$, $\delta(\mathcal H)$ is the sentence
\[
\exists x_1\ldots\exists x_n\big(\bigwedge_{1\le i<j\le n}x_i\neq x_j\wedge
\bigwedge_{\substack{1\le i,j\le n\\E^{\mathcal H}h_ih_j}}Ex_ix_j\big).
\]
Slicewise $\Phi$-definability is closely related to the model-checking
problem for $\Phi$: If $P\subseteq\STR\times\Pi^*$ is slicewise
$\Phi$-definable, then $P\prm\MC(\Phi)$.  On the
other hand, 
if $\Phi$ is a decidable set of formulas, then the problem
$\MC(\Phi)$ is slicewise $\Phi$-definable for trivial reasons.

\begin{Definition}\label{def:4}
  Let $\tau$ be a vocabulary, $P\subseteq\STR[\tau]\times\mathbb N$ a
  parameterized problem, and $\Phi$ a class of formulas. $P$ is
  \emph{$\Phi$-Fagin-definable} if there is a relation symbol $X\not\in\tau$
  (say, $r$-ary) and a sentence $\phi\in\Phi[\tau\cup\{X\}]$ such that for 
all
  $\mathcal A\in\STR[\tau]$ and $k\in\mathbb N$ we have $(\mathcal A,k)\in 
P$
  if and only if there is a $B\subseteq A^r$ such that $|B|=k$ and $(\mathcal
  A,B)\models\phi$. (Here $(\mathcal A,B)$ denotes the
  $\tau\cup\{X\}$-expansion of $\mathcal A$ that interprets $X$ by 
  $B$.) Then $\phi$ \emph{Fagin-defines} $P$.
\end{Definition}

\noindent
We often consider $X$ as a relation \emph{variable} and thus write 
$\phi(X)\in\Phi[\tau]$ instead of $\phi\in\Phi[\tau\cup\{X\}]$ and $\mathcal
  A\models\phi(B)$ instead of $(\mathcal
  A,B)\models\phi$.

For example, parameterized vertex cover \VC\ is Fagin-defined by the formula
\[
\phi_{\VC}:=\forall y\forall z\big(Eyz\rightarrow (Xy\vee Xz)\big).
\]
It is easy to see that every problem that is \FO-Fagin-definable is also
\FO-slicewise definable. Indeed, if $P\subseteq\STR[\tau]\times\mathbb N$ is
Fagin-defined by a formula $\phi(X)\in\FO[\tau]$, where $X$ is $r$-ary,
then it is slicewise \FO-defined via the function $\delta:\mathbb
N\rightarrow\FO[\tau]$ with $\delta(k)=\exists\bar x_1\ldots\exists\bar
x_k(\bigwedge_{1\le i<j\le k}\bar x_i\neq \bar x_j \wedge
\phi_k)$, where $\bar x_1,\ldots,\bar x_k$ are $r$-tuples of distinct
variables not occurring in $\phi$ and $\phi_k$ is the sentence obtained from
$\phi$ by replacing each subformula of the form $X\bar y$ by
$\bigvee_{i=1}^k\bar x_i=\bar y$.
Fagin-definability implies slicewise definability also for other 
reasonable classes $\Phi$ of formulas, e.g., for the class 
$\Sigma^1_{1}$ of 
formulas of second-order logic of the form $\exists X_{1}\ldots 
\exists X_{l}\psi$, where $X_{1},\ldots X_{l}$ are relation variables 
and $\psi$ is first-order.

The converse is certainly not true, not even for problems of the specific form
$P\subseteq\STR[\tau]\times\mathbb N$: It is obvious that 
there are slicewise \FO-definable problems of arbitrarily
high classical complexity (choose $\delta$ in Definition \ref{def:3} arbitrarily
complex). On
 the other hand we have the following 
characterization of $\Sigma^1_{1}$-Fagin-definable
problems.

\begin{Proposition}\label{theo:1}
Let $P\subseteq\STR[\tau]\times\mathbb N$. Then, (1) and (2) are 
equivalent, where
\begin{enumerate}
\item $P$ is $\Sigma^1_{1}$-Fagin-definable.
\item $P$ is in {\rm NP} (when considered as a classical 
problem) and for some $r\ge 1$, $(\mathcal A,k)\in P$ implies $k\le
|A|^r$.
\end{enumerate}
\end{Proposition}

\proof
The implication of (1) $\Rightarrow$ (2) being clear, we turn to a 
proof of (2) $\Rightarrow$ (1). Choose $r$ according to (2). Then, 
$$
\{(\mathcal A,B) \mid B\subseteq A^r \mbox{ and } (\mathcal A,|B|)\in 
P\}
$$
is a class of $\tau\cup \{X\}$-structures in NP, where $X$ is $r$-ary.
By Fagin's Theorem \cite{fag74}, there is a 
$\Sigma^1_{1}$-formula $\varphi(X)$ of vocabulary $\tau\cup \{X\}$ 
axiomatizing this class. Then, $\varphi(X)$  
$\Sigma^1_{1}$-Fagin-defines $P$.
\proofend

 Thus, slicewise definability is the more
general notion. Nevertheless, Fagin definability can be very
useful. We illustrate this by the following  generalization of a
result due to Cai and Chen, namely Theorem 3.5 of \cite{caiche97}, which is based on a
result due to Kolaitis and Thakur \cite{koltha95} that syntactically
characterizes certain minimization problems. It is motivated by
comparing the formula $\phi_{\VC}$ defining the fixed-parameter
tractable problem \VC\ with the following formulas $\phi_{\DS}$ and
$\phi_{\CLI}$ defining the $\W[1]$-hard problems $\DS$ and $\CLI$,
respectively:
\begin{align*}
\phi_\DS&:=\forall y\exists x(Xx\wedge(x=y\vee Exy)\big),\\
\phi_\CLI&:=\forall y\forall z\big((Xy\wedge Xz)\rightarrow (y=z\vee Eyz)\big).
\end{align*}
Observe that in $\phi_{\DS}$ the relation variable $X$ is in the scope
of an existential quantifier and in $\phi_{\CLI}$ it occurs negatively.

\begin{Theorem}\label{theo:2}
  Let $\tau$ be a vocabulary and $P\subseteq \STR[\tau]\times\mathbb N$ a
  parameterized problem that is Fagin-defined by a
  $\FO[\tau]$-formula $\phi(X)$ in which $X$ does not occur in the
  scope of an existential quantifier or negation symbol. Then $P$ is in \FPT.
\end{Theorem}

\proof For simplicity, let us assume that $X$ is unary.  Without loss
of generality we can assume that 
\[
\phi=\forall y_1\ldots\forall
y_l\bigvee_{i=1}^m\bigwedge_{j=1}^{p}\psi_{ij}, 
\]
where each
$\psi_{ij}$ either is $Xy_q$ for some $q\in\{1,\ldots,l\}$, or a
first-order formula with free variables in $\{y_1,\ldots,y_l\}$ in
which $X$ does not occur.  In a preprocessing phase we replace the
latter ones by atomic formulas: For each such $\psi_{ij}$ we introduce
a new relation symbol $R_{ij}$ whose arity matches the number of free
variables of $\psi_{ij}$ and let $\tau^*$ be the set of all these
relation symbols. We let $\phi^*$ be the formula obtained from $\phi$
by replacing each subformula $\psi_{ij}(\bar z)$ by $R_{ij}\bar z$.
Then $\phi^*=\forall y_1\ldots\forall y_l\bigvee_{i=1}^m\chi_i$, where
each $\chi_i$ is a conjunction of atomic formulas. 

For a structure $\mathcal A\in\STR[\tau]$ we let $\mathcal A^*$ be
the $\tau^*$-structure with universe $A$ and  with
\[
R_{ij}^{\mathcal A^*}:=\{\bar a\mid\mathcal A\models\psi_{ij}(\bar
a)\}.
\]
Then we have for $B\subseteq A$,
\[
\mathcal A\models\phi(B)\iff\mathcal A^*\models\phi^*(B).
\]
Given $\mathcal A$, each $R_{ij}^{\mathcal A^*}$ can be computed in
time $O(|A|^{||\psi_{ij}||})$, thus $\mathcal A^*$ can certainly be computed in
time $O(|A|^{||\phi||})$.

For $1\le i\le m$, $\bar{a}=a_1\ldots a_l\in A^l$, and $B\subseteq A$ we 
let
\[
\beta(B,\bar{a},i):=B \cup \{a_j\mid Xy_j\text{ is a conjunct of
  }\chi_i\}.
\]
Since $\chi_i(X, \bar{y})$ is positive in $X$, the following two
statements are equivalent for every $B'$ with $B\subseteq B' \subseteq
A$:

\begin{itemize}
\item ${\mathcal A}^*\models \chi_i(B',\bar{a})$.
\item $\beta(B,\bar a,i)\subseteq B'$ and ${\mathcal
    A}^{*}\models\chi_i(\beta(B,\bar{a},i),\bar{a})$.
\end{itemize}
This equivalence is used by Algorithm \ref{alg:1}
to decide $P$.

\begin{figure}
\begin{center}
  \fbox{\parbox{11cm}{\setcounter{zeilencounter}{0}
      \textsc{Check-$\phi$}($\mathcal A\in\STR[\tau]$, $k\in\mathbb N$)
\begin{tabbing}
\hspace*{4mm}\=\hspace{9mm}\=\hspace{6mm}\=\hspace{6mm}\=\hspace{6mm}\=\hspace{6mm}\=\hspace{6mm}\=\hspace{6mm}\=\kill
\zeile compute $\mathcal A^*$\\
\zeile initialize set $\mathcal S\subseteq\Pow(A)$ by
$\mathcal S:=\{\emptyset\}$\\
\zeile \FORALL\ $\bar a\in A^l$ \DO\\
\zeile\>\FORALL\ $B\in\mathcal S$ \DO\\
\zeile\>\>\IF\ $\mathcal A^*\not\models\bigvee_{i=1}^m\chi_i(B, \bar a)$ \THEN\\
\zeile\>\>\>$\mathcal S:=\mathcal S\setminus\{B\}$\\
\zeile\>\>\>\FOR\ $i=1$ \TO\ $m$ \DO\\
\zeile\>\>\>\>compute $B':=\beta(B,\bar a,i)$\\
\zeile\>\>\>\>\IF\ $|B'|\le k$ \AND\ $\mathcal
A^*\models\chi_i(B',\bar a)$\\
\zeile\>\>\>\>\>\THEN\ $\mathcal S:=\mathcal S\cup\{B'\}$\\
\zeile\IF\ $S\neq\emptyset$\\
\zeile\>\THEN\ \ACCEPT\\
\zeile\>\ELSE\ \REJECT.
\end{tabbing}}}
\end{center}
\alg\label{alg:1}
\end{figure}

Recall that, given a $\tau$-structure $\mathcal A$ and a parameter
$k\in\mathbb N$, the algorithm is supposed to decide whether there is
a $B\subseteq A$ with $|B|=k$ such that for all $\bar a\in A^l$ there
is an $i$ such that $\mathcal A^*\models\chi_i(B,\bar a)$.

The crucial observation to see that the algorithm is correct is that whenever
the main loop in Lines 3--10 is entered, $\mathcal S$ is a set of subsets
$B\subseteq A$ such that $|B|\le k$ and for all $\bar a$ considered so far (in
earlier runs through the loop) we have $\mathcal
A^*\models\bigvee_{i=1}^m\chi_i(B,\bar a)$.

To get a bound on the running time, we note that whenever a new set is added
to $\mathcal S$ (in Line 10) then it is an extension of a strictly smaller set
that has just been removed from $\mathcal S$ (in Line 6). Furthermore, for
each set removed (in Line 6) at most $m$ such extensions can be added. Thus an
upper bound for the number of sets  that can be in $\mathcal S$ at any time is
$m^k$. The main loop (in Lines 3--10) is called $n^l$ times, where $n:=|A|$.
This gives an overall bound on the running time of $O(m^k\cdot n^l)$ plus the time
needed to compute $\mathcal A^*$.

Since $l$ does not depend on the instance, but just on the formula $\phi$,  this yields 
the fixed-parameter tractability of $P$.
\proofend

Besides \VC, many other parameterized problems can be shown to be
fixed-parameter tractable by a simple application of this theorem.
Let us consider one example in detail:
The
\emph{valence} of a graph is the maximal number of neighbors a vertex in the
graph has. For an $l\ge 1$, we consider the restriction of \textit{DOMINATING
  SET} to graphs of valence at most $l$, i.e.\ the problem

\PP{VC$_l$}{Graph $\mathcal G$}{$k\in\mathbb N$}{Is the valence of $\mathcal
  G$ at most $l$ and does $\mathcal G$ have a dominating set of size
  at most $k$}

\noindent
This problem is Fagin-defined by the following first-order formula:
\[
\forall x\exists^{\le l}z\;Exz\wedge\forall y_0\forall
y_1\ldots\forall y_l\big(\forall
z(Ey_0z\rightarrow\bigvee_{i=1}^lz=y_i)\rightarrow\bigvee_{i=0}^lXy_i\big).
\]
($\exists^{\le m}x\psi(x)$ abbreviates $\exists y_1\ldots\exists y_m\forall
x(\psi(x)\rightarrow\bigvee_{i=1}^m x=y_i)$.)

Other examples of problems that can be shown to be in \FPT\ by Theorem
\ref{theo:1} are \textit{HITTING SET FOR SIZE THREE SETS}, \textit{MATRIX 
DOMINATION}, or \textit{SHORT 3DIMENSIONAL MATCHING} of
\cite{dowfel99}.

\section{Homomorphisms, embeddings, and model-checking}\label{sec:5}
In this section we analyze the  close relationship between the
homomorphism problem, the embedding problem, and model-checking
problems for $\Sigma_1$-formulas from the point of view of parameterized complexity (compare \cite{kolvar98} for a further analysis of this relationship).

A \emph{homomorphism} from a $\tau$-structure $\mathcal B$ into a
$\tau$-structure $\mathcal A$ is a mapping $h:B\rightarrow A$ such
that for all $R\in\tau$ and tuples $\bar b\in R^{\mathcal B}$ we have
$h(\bar b)\in R^{\mathcal A}$.
The parameterized \textit{HOMOMORPHISM PROBLEM (HOM)} is defined as follows:

\PP{HOM}{$\mathcal A\in\STR$}{$\mathcal B\in\STR$}{Is there a homomorphism
  from $\mathcal B$ to $\mathcal A$}

\noindent
A \emph{(weak) embedding} of $\mathcal B$ into $\mathcal A$ is an
injective homomorphism from $\mathcal B$ to $\mathcal A$. Note that a
graph $\mathcal H$ is isomorphic to a subgraph of a graph $\mathcal G$
in the usual graph theoretic sense, if there is an embedding of
$\mathcal H$ into $\mathcal G$. Thus the following parameterized
\textit{EMBEDDING PROBLEM (EMB)} is a generalization of the subgraph
isomorphism problem \textit{SI}:

\PP{EMB}{$\mathcal A\in\STR$}{$\mathcal B\in\STR$}{Is there an embedding
  of $\mathcal B$ into $\mathcal A$}

\noindent\label{page:gaif} 
The \emph{Gaifman graph} of a $\tau$-structure
$\mathcal A$ is the graph $\mathcal G(\mathcal A)$ with universe $A$
in which two elements $a\neq b$ are adjacent if there is an $R\in\tau$
and a tuple $\bar a\in R^{\mathcal A}$ such that both $a$ and $b$
occur in the tuple $\bar a$. For a class $C$ of graphs we let
$\STR[C]$ denote the class of all structures whose Gaifman graph is in
$C$. Note that $\STR[\GRA\,]=\STR$. Furthermore, we let
$\STR[\tau,C]:=\STR[\tau]\cap\STR[C]$ for every vocabulary $\tau$ and
$\STR[s,C]:=\STR[s]\cap\STR[C]$ for every $s\ge 1$.  We define
restrictions $\HOM[\ldots]$ and $\EMB[\ldots]$ of the respective
problems, where for every possible restriction `$\ldots$' in the
square brackets we require the \emph{parameter} $\mathcal B$ to belong to the
class $\STR[\ldots]$. For example, we let
\[
\EMB[s,C]:=\EMB\cap(\STR\times\STR[s,C]).
\]

\begin{Lemma}\label{lem:3}
For all classes $C$ of graphs and $s\ge 1$ we have
$\HOM[C]\pprm\EMB[C]$ and $\HOM[s,C]\pprm\EMB[s,C]$.
\end{Lemma}

\proof Suppose we are given an instance $(\mathcal A,\mathcal B)$ of
$\HOM[C]$. Let $\tau$ be the vocabulary of $\mathcal A$. Let $\mathcal
A_B$ be the $\tau$-structure which for every element of $\mathcal A$
contains $|B|$ duplicates, i.e.\ $A_B:=A\times B$ and, for
every $r$-ary  $R\in\tau$,
\begin{align*}
R^{\mathcal A_B}&:=\big\{((a_1,b_1),\ldots ,(a_r,b_r))\bigmid
R^{\mathcal A}a_1\ldots a_r\}.
\end{align*}
Then, every homomorphism $h:\mathcal B\rightarrow\mathcal A$ gives rise
to an embedding $f:\mathcal B\rightarrow\mathcal A_B$ defined by
$f(b)=(h(b),b)$ and every embedding $f:\mathcal B\rightarrow\mathcal
A_B$ induces a homomorphism $h:\mathcal B\rightarrow\mathcal A$
defined by letting $h(b)$ be the projection on the first component of $f(b)$.
\proofend

Note that, unless $\PTIME=\NP$, there is no polynomial-time reduction from
$\EMB[C]$ to $\HOM[C]$ for the class $C$ of all paths, because $\HOM[C]$ can
easily be seen to be in $\PTIME$ by a dynamic programming algorithm, whereas
$\EMB[2,C]$ is NP-complete by a reduction from \textit{HAMILTONIAN PATH}.
Thus $\EMB[s,C]\pprm\HOM[s,C]$ does not hold for any $s\ge 2$. However, we
will see that for all $s\ge 2$ and $C$ we actually have
$\HOM[s,C]\eprT\EMB[s,C]$.

Before we show this, we introduce two related model-checking problems.
With each first-order formula $\phi$ we associate a graph $\mathcal
G(\phi)$. Its universe is $\var(\phi)$, the set of all variables in
$\phi$, and there is an edge between distinct
$x,y\in\var(\phi)$ in $\mathcal G(\phi)$ if $\phi$ has an atomic
subformula in which both $x,y$ occur. 

Let $\phi^{\neq}$ be the formula obtained from $\phi$ by deleting all
inequalities, i.e.\ all atomic subformulas of the form $x=y$ that
occur in the scope of an odd number of negation symbols. We are also
interested in $\mathcal G(\phi^{\neq})$. Let us see an example:
\[
\phi:=\exists x_1\ldots\exists x_k\big(\bigwedge_{1\le i<j\le k}\neg
x_i=x_j\wedge \bigwedge_{i=1}^{k-1}Ex_ix_{i+1}\big).
\]
$\mathcal G(\phi)$ is the complete graph with
vertex set $\{x_1,\ldots,x_k\}$, whereas $\mathcal G(\phi^{\neq})$ is
the path $x_1\ldots x_k$. Note that $\phi$ says that a graph
has a subgraph isomorphic to a path of length $k$, whereas
$\phi^{\neq}$ says that a graph contains a
homomorphic image of a path of length $k$. This generalizes to the
following simple lemma, whose proof we omit.

\begin{Lemma}\label{lem:4}
  For every structure $\mathcal B\in\STR$ there is a
  $\Sigma_1$-sentence $\phi_{\mathcal B}$ (whose quantifier-free part
  is a conjunction of literals) such that $\mathcal G(\phi_{\mathcal
    B}^{\neq})=\mathcal G(\mathcal B)$, and for every structure
  $\mathcal A$ we have:
  \begin{align*}
    \mathcal A\models\phi_{\mathcal B}&\iff\text{There is an embedding
    of $\mathcal B$ into $\mathcal A$.}\\
    \mathcal A\models\phi_{\mathcal B}^{\neq}&\iff\text{There is a homomorphism from
    $\mathcal B$ to $\mathcal A$.}
  \end{align*}
  Furthermore, the mapping $\mathcal B\mapsto\phi_{\mathcal B}$ is computable in linear time.
\end{Lemma}

For $s\in\mathbb N$ and a class $C$ of graphs we let
\begin{align*}
\Sigma_1[s,C]&:=\big\{\phi\in\Sigma_1[\tau]\bigmid\tau\;s\text{-ary
  vocabulary},\mathcal G(\phi)\in C\big\},\\
\Sigma_1^{\neq}[s,C]&:=\big\{\phi\in\Sigma_1[\tau]\bigmid\tau\;s\text{-ary
  vocabulary},\mathcal G(\phi^{\neq})\in C\big\}.
\end{align*}
Furthermore, we let $\Sigma_1[C]:=\bigcup_{s\ge 1}\Sigma_1[s,C]$ and $\Sigma_1^{\neq}[C]:=\bigcup_{s\ge 1}\Sigma_1^{\neq}[s,C]$.

Lemma \ref{lem:4} implies that for every $s\ge 1$ and for every
class $C$ of graphs we have $\HOM[s,C]\pprm\MC(\Sigma_1[s,C])$ and
$\EMB[s,C]\pprm\MC(\Sigma_1^{\neq}[s,C])$.
Unless $\PTIME=\NP$, the converse of
these statements is wrong. To see this, let $C$ be the class of all
graphs that only consist of isolated vertices, i.e.\ all graphs $\mathcal G$ with $E^{\mathcal G}=\emptyset$. Then
clearly $\HOM[C]$ and $\EMB[C]$ are in $\PTIME$, but we can reduce the
satisfiability problem for propositional formulas to $\MC(\Sigma_1[1,C])$
and $\MC(\Sigma_1^{\neq}[1,C])$.

\begin{Theorem}\label{theo:3}
  Let $C$ be a class of graphs and $s\ge 2$. Then
\[
\HOM[s,C]\eprT\EMB[s,C]\eprT
\MC(\Sigma_1^{\neq}[s,C])\eprT\MC(\Sigma_1[s,C]).
\]
\end{Theorem}

\proof We have already seen that $\HOM[s,C]\prT\EMB[s,C]$ (Lemma
\ref{lem:3}) and that
$\EMB[s,C]\prT\MC(\Sigma_1^{\neq}[s,C])$ (Lemma \ref{lem:4}). To complete the cycle
we shall prove that $\MC(\Sigma_1^{\neq}[s,C])\prT\MC(\Sigma_1[s,C])$
and that $\MC(\Sigma_1[s,C])\prT\HOM[s,C]$.

\medskip 
We first prove that $\MC(\Sigma_1[s,C])\prT\HOM[s,C]$. Let
$\phi\in\Sigma_1[s,C]$, say, of vocabulary $\tau$, and $\mathcal A$ a
$\tau$-structure. We shall describe an algorithm that decides whether
$\mathcal A\models\phi$ using $\HOM[s,C]$ as an oracle.

Let $S$ be a binary relation symbol not contained in $\tau$ and
$\tau':=\tau\cup\{S\}$. Furthermore, let $\mathcal A'$ be the
$\tau'$-expansion of $\mathcal A$ with $S^{\mathcal A'}=\emptyset$.
Our algorithm first computes a sentence
$\phi':=\bigvee_{i=1}^m\exists\bar x_i\psi_i$ of vocabulary $\tau'$ such that
\begin{enumerate}
\item
$\mathcal A\models\phi$ if, and only if, $\mathcal A'\models\phi'$.
\item
For $1\le i\le m$, the formula $\psi_i$ is a conjunction of literals,
and we have $\mathcal G(\psi_i)=\mathcal G(\phi)$.
\end{enumerate}
This can be achieved by first translating $\phi$ to a sentence whose
quantifier-free part is in disjunctive normal form, then swapping
existential quantifiers and the disjunction, and then adding dummy
literals of the form $\neg Sxy$ until $\mathcal G(\psi_i)=\mathcal G(\phi)$.

Let $\tau'':=\tau'\cup\{\overline R\mid
R\in\tau'\}\cup\{E,\overline E\}$, where for all $R\in\tau'$ the symbol $\overline
R$ is a new relation symbol of the same arity as $R$ and $E,\overline E$
are new binary relation symbols. Let $\mathcal A''$ be the $\tau''$
expansion of $\mathcal A'$ in  which $\overline R$ is interpreted as the
complement of $R^{\mathcal A''}$ and $E,\overline E$ are interpreted as
equality and inequality, respectively. 
For $1\le i\le m$, we define a
$\tau''$-structure $\mathcal B_i$ with 
$\mathcal G(\mathcal B_i)=\mathcal G(\phi)$
 such that $\mathcal
A'\models\exists\bar x_i\psi_i$ if, and only if, there is a
homomorphism from $\mathcal B_i$ into $\mathcal A''$. 
We let $\mathcal B_i$ be the $\tau''$-structure with universe
$\var(\psi_i)$ and 
\begin{align*}
R^{\mathcal B_i}&:=\{\overline y\mid R\overline y\text{ is a literal of
  }\psi_i\}\qquad(\text{for }R\in\tau'),\\
\overline R^{\mathcal B_i}&:=\{\overline y\mid \neg R\overline y\text{ is a literal of
  }\psi_i\}\qquad(\text{for }R\in\tau'),\\
E^{\mathcal B_i}&:=\{yz\mid y=z\text{ is a literal of
  }\psi_i\},\\
\overline E^{\mathcal B_i}&:=\{yz\mid \neg y=z\text{ is a literal of
  }\psi_i\}.
\end{align*}
It is obvious that $\mathcal B_i$ does indeed have the desired
property.  Altogether, our construction yields a parameterized T-reduction.

\medskip 
It remains to prove that
$\MC(\Sigma_1^{\neq}[s,C])\prT\MC(\Sigma_1[s,C])$. We use the so called \emph{color coding} technique
of Alon, Yuster, and Zwick \cite{aloyuszwi95}.

Let $l\ge 1$ and $X$ a set. An \emph{$l$-perfect family of hash functions on
  $X$} is a family $F$ of functions $f:X\rightarrow\{1,\ldots,l\}$ such that
for all subsets $Y\subseteq X$ of size $l$ there is an $f\in F$ such that
$f(Y)=\{1,\ldots,l\}$ (i.e.\ on $Y$, $f$ is one-to-one).  Alon, Yuster, and Zwick \cite{aloyuszwi95} show that
given $n,l\ge 1$, an $l$-perfect family of hash functions on $\{1,\ldots,n\}$
of size $2^{O(l)}\cdot\log n$ can be computed in time $2^{O(l)}\cdot n\cdot\log n$.

For a similar reason as outlined above, without loss of generality we can
restrict our attention to sentences $\phi\in\MC(\Sigma_1^{\neq}[s,C])$
whose quantifier-free part is a conjunction of literals. Given such a
sentence $\phi=\exists x_1\ldots\exists x_k\psi$, say of vocabulary $\tau$, and a $\tau$-structure
$\mathcal A$, we define a family of  sentences
$\phi_\gamma\in\MC(\Sigma_1[s,C])$ and a family of structures $\mathcal
A_f$ such that $\mathcal A\models\phi$ if, and
only if, there is a $\gamma$ and $f$ such that $\mathcal
A_f\models\phi_\gamma$.

A \emph{coloring of} $\phi$ is a function $\gamma:\{1,\ldots,k\}\rightarrow
\{1,\ldots,k\}$ such that $\gamma(i)\not=\gamma(j)$ if $\neg x_i=x_j$ occurs
in $\phi$. For a coloring $\gamma$ we let $\psi_{\gamma} $ be the formula
obtained from $\psi$ by replacing all literals $\neg x_i=x_j$ by
$C_{\gamma(i)}x_i\wedge C_{\gamma(j)}x_j$ (here, $C_1,\ldots,C_k$ are new unary
``color" relation symbols) and let $\phi_\gamma:=\exists x_1\ldots\exists x_k\psi_\gamma$.
Note that $\mathcal G(\phi_{\gamma})=\mathcal G(\phi^{\neq})$. Thus
$\phi_\gamma\in\MC(\Sigma_1[s,C])$.

With every  $f:A\rightarrow\{1,\ldots,k\}$, which we call a
\emph{coloring of $\mathcal A$}, we let $\mathcal A_f$ be the
$\tau\cup\{C_1,\ldots,C_k\}$-expansion of $\mathcal A$ with
$C_i^{\mathcal A_f}:=f^{-1}(i)$ for $1\le i\le k$.

Observe that 
\begin{equation}\label{eq:cc1}
{\mathcal A}\models \phi\iff\parbox[t]{8cm}{there is a coloring
  $\gamma$ of $\phi$ and a coloring $f$ of $\mathcal A$ such that ${\mathcal
    A}_f\models\phi_{\gamma}$.}
\end{equation}
The problem is that there are $k^{|A|}$ colorings of $\mathcal A$, so
\eqref{eq:cc1} does not yet give rise to a parameterized reduction.
The crucial trick is that to achieve this equivalence we do not
have to consider all possible colorings $f$ of $\mathcal A$. For $1\le
l\le k$, let $F_l$ be an $l$-perfect family of hash-function on $A$
and $F:=\bigcup_{l=1}^kF_l$. We claim that
\begin{equation}\label{eq:cc2}
{\mathcal A}\models \phi\iff\parbox[t]{9cm}{there is a coloring
  $\gamma$ of $\phi$ and an $f\in F$ such that ${\mathcal
    A}_f\models\phi_{\gamma}$.}
\end{equation}
The backward direction follows immediately from \eqref{eq:cc1}. For
the forward direction, suppose that $\mathcal A\models\phi$.
 Let $\bar a\in A^k$ such that $\mathcal
A\models\psi(\bar a)$. 
There is a function $f\in F$ whose restriction to $\{a_1,\ldots,a_k\}$ is one-to-one. Define $\gamma$ by $\gamma(i):=f(a_i)$. Then, $\gamma$ is a coloring of
$\phi$, ${\mathcal A}_f\models\psi_{\gamma}(\bar a)$ and hence, 
${\mathcal A}_f\models\phi_{\gamma}$.

Since the family $F$ can be chosen sufficiently small and computed
sufficiently fast, the equivalence \eqref{eq:cc2} gives rise to a
parameterized reduction.
\proofend

\begin{Remark}
  We do not know if the parameterized T-reductions in Theorem \ref{theo:3}
  can be replaced by parameterized m-reductions. However, for many
  interesting classes $C$ they can be replaced. One such example is
  the class of all graphs.
  Similar techniques work for all classes $C$
  of graphs for which there exists an algorithm that, given a graph
  $\mathcal H\in C$, computes a connected $\mathcal H'\in C$ such that
  $\mathcal H$ is a subgraph of $\mathcal H'$. For all such classes $C$ we can
  show that
\[
\HOM[s,C]\eprm\EMB[s,C]\eprm
\MC(\Sigma_1^{\neq}[s,C])\eprm\MC(\Sigma_1[s,C]).
\]
(for all $s\ge 1$).
\end{Remark}

\subsection{Sentences of bounded tree-width}
Our main application of Theorem \ref{theo:3} is to sentences whose underlying
graphs have bounded tree-width. In the time since we submitted this article,
considerable progress has been made in this area.  For an extensive discussion
of model-checking algorithms based on tree-decompositions of the sentences, we
refer the reader to \cite{flufrigro01}.

\medskip
We think of a \emph{tree} $\mathcal T$ as directed
from its root, which we denote by $r^{\mathcal T}$, to the leaves and thus can speak of a \emph{child} and of the
\emph{parent} of a vertex.

A \emph{tree-decomposition} of a $\tau$-structure ${\mathcal A}$ is a pair
$(\mathcal T,(A_t)_{t\in T})$, where $\mathcal T$ is a tree and $(A_t)_{t\in
  T}$ a family of subsets of $A$ such that
\begin{enumerate}
\item For every $a\in A$, the set $\{t\in T\mid a\in A_t \}$
  is non-empty and induces a subtree of $\mathcal T$ (that is, is connected).
\item For every $k$-ary relation symbol $R\in\tau$ and all $a_1,\ldots,a_k\in
  A$ such that $R^{\mathcal A}a_1,\ldots,a_k$ there exists a $t\in T$ such
  that $a_1,\ldots,a_k\in A_t$. 
\end{enumerate}
The \emph{width} of a tree-decomposition $(\mathcal T,(A_t)_{t\in T})$ is $\max \{|A_t|
\mid t\in T \}-1$. The \emph{tree-width} $\tw({\mathcal A})$ of  ${\mathcal A}$ is the minimal width of a tree-decomposition of
${\mathcal A}$.

For $s\ge 1$, let $W_s$ denote the class of all structures of tree-width at
most $s$ and $\GW_s:=\GRA\cap W_s$. Note that $W_s\subseteq\STR[s+1]$, because
a graph of tree-width $s$ has clique number at most $s+1$.\footnote{This is
  slightly imprecise, because a structure $\mathcal A\in W_s$ might have a
  vocabulary of arbitrarily high arity, as long as no tuple contained in a
  relation of $\mathcal A$ consists of more than $s+1$ distinct elements. But
  since any structure with this property can easily be transformed to an
  $(s+1)$-ary structure that is essentially the same, we decided to accept
  this imprecision in exchange for a simpler
  notation.}\addtocounter{footnote}{-1}\refstepcounter{footnote}\label{fuss}
(The clique number of a graph $\mathcal G$ is the maximal cardinality of a set
of pairwise adjacent vertices of $\mathcal G$.)
 
Plehn and Voigt \cite{plevoi90} were the first to realize that tree-width is
a relevant parameter for the problems considered in the previous section. They
proved that the parameterized embedding problem restricted to (parameter)
graphs of bounded tree-width is fixed parameter tractable. Chekuri and
Rajaraman \cite{cheraj97} proved that for every $s\ge 1$ the problem
$\HOM[GW_s]$ is in PTIME (when considered as an unparameterized problem)
and therefore fixed-parameter tractable. They phrased their result in terms
of the equivalent \emph{conjunctive query containment} problem (also see \cite{kolvar98}).

Thus as a corollary of Theorem \ref{theo:3} we obtain:

\begin{Corollary}\label{cor:4}
  Let $s\ge 1$. Then the problems $\EMB[\GW_s]$,
  $\MC(\Sigma_1[\GW_s])$, and $\MC(\Sigma_1^{\neq}[\GW_s])$ are
  in \FPT.
\end{Corollary}

Papadimitriou and Yannakakis \cite{papyan97} proved the model-checking results
of this corollary for the related case of acyclic conjunctive queries.

Unfortunately, it turns out that Corollary \ref{cor:4} is the only real
application of Theorem \ref{theo:3}. Very recently, Schwentick, Segoufin, and
the second author \cite{groschweseg01} have proved that for every class $C$ of
graphs of unbounded tree-width and every $s\ge 2$, the problem $\HOM[C,s]$ is
$\W[1]$-complete under parameterized T-reductions.

\section{\FO-model-checking on graphs with excluded minors}
The fixed-parameter tractability results of the previous section were
obtained by putting syntactical restrictions on the sentences,
i.e.\ the \emph{parameter} of the
model-checking problem.
In this section we put restrictions on the structures, i.e.\ the
\emph{input} of the model-checking problem.

Recall the definition of the parameterized problem $\MC(\Phi)|_D$, for
a class $\Phi$ of formulas and a class $D$ of structures:

\PP{$\MC(\Phi)|_D$}{$\mathcal A\in\STR$}{$\phi\in\Phi$}{Is $\mathcal A\in
 D$ and $\mathcal A\models\phi$}

\noindent
Our starting point is the following theorem due to Courcelle. Remember
that \emph{monadic second-order logic} is the extension of first-order
logic where one is allowed to quantify not only over individual
elements of a structure but also over sets of elements. \MSO\ denotes
the class of all formulas of monadic second-order logic. Remember
that $W_s$ denotes the class of all structures of
tree-width at most $s$ (for $s\ge 0$).

\begin{Theorem}[\cite{cou90}]\label{theo:7}
Let $s\ge 0$. Then $\MC(\MSO)|_{W_s}$ is in \FPT.
\end{Theorem}

A graph $\mathcal H$ is a \emph{minor} of a graph $\mathcal G$ (we
write $\mathcal H\preceq \mathcal G$) if $\mathcal H$ is can be
obtained from a subgraph of $\mathcal G$ by contracting edges.
$\mathcal H$ is an \emph{excluded minor} for a class $D$ if $\mathcal
H$ is not a minor of any graph in $D$. Note that a class $D$ of graphs
has an excluded minor if, and only if, there is an $n\in\mathbb N$
such that $\mathcal K_n$ is an excluded minor for $D$.

Examples of classes of graphs with an excluded minor are classes of graphs
of bounded tree-width or classes of graphs embeddable in a fixed
surface. 

Recall that for a class $D$ of graphs, $\STR[D]$ denotes the class of
all structures whose Gaifman graph is in $C$.

\begin{Theorem}\label{theo:8}
  Let $D$ be a \PTIME-decidable class of graphs with an excluded minor. Then
  $\MC(\FO)|_{\STR[D]}$ is in $\FPT$.
\end{Theorem}

The rest of this section is devoted to the proof of this theorem,
which needs some preparation.

A class $D$ of graphs is called \emph{minor closed} if for all
$\mathcal G\in D$ and $\mathcal H\preceq\mathcal G$ we have $\mathcal
H\in D$.  Robertson and Seymour proved:

\begin{Theorem}[\cite{GMXIII}]\label{theo:9}
  Every minor-closed class of graphs is \PTIME-decidable.
\end{Theorem}

This, together with Theorem \ref{theo:8}, immediately yields:

\begin{Corollary}\label{cor:5}
  Let $D\varsubsetneqq GRAPH$ be minor closed. Then
  $\MC(\FO)|_{\STR[D]}$ is in $\FPT$.
\end{Corollary}

 Recall the
definition of the Gaifman graph $\mathcal G(\mathcal A)$ of a
structure $\mathcal A$ (cf.\ Page \pageref{page:gaif}).  The distance
$d^{\mathcal A}(a,b)$ between $a,b\in A$ is the
length of the shortest path from $a$ to $b$ in $\mathcal G({\mathcal
  A})$. For $r\in\mathbb N$ and $a\in A$, the \emph{$r$-ball} around
$a$ is the set $B_r^{\mathcal A}(a):=\{b\in A\mid d^{\mathcal
  A}(a,b)\le r\}$. For an $X\subseteq A$, $\langle X\rangle^{\mathcal
  A}$ denotes the substructure induced by $\mathcal A$ on $X$, i.e.\ 
the structure with universe $X$ and $R^{\langle X\rangle^{\mathcal
    A}}=R^{\mathcal A}\cap X^r$ for all $r$-ary relation symbols $R$
in the vocabulary of $\mathcal A$. Furthermore, we let $\mathcal
A\setminus X:=\langle A\setminus X\rangle^{\mathcal A}$.

The \emph{local tree-width} of $\mathcal A$ is the
function $\ltw(\mathcal A):\mathbb N\rightarrow\mathbb N$ defined by
\[
 \ltw(\mathcal A)(r):=\max\big\{\tw(\langle B_r^{\mathcal
 A}(a)\rangle^{\mathcal A})\bigmid a\in A\big\}.
\]
For functions $f,g:\mathbb N\rightarrow\mathbb N$ we write $f\le g$ if
$f(n)\le g(n)$ for all $n\in\mathbb N$.  A class $D$ of structures has
\emph{bounded local tree-width} if there is a function $\lambda:\mathbb
N\rightarrow\mathbb N$ such that for all $\mathcal A\in D$ we have
$\ltw(\mathcal A)\le\lambda$.

The ``local'' character of first-order formulas allows to generalize Theorem \ref{theo:7} for
first-order logic from classes of structures of bounded tree-width to 
classes of structures of bounded local tree-width:

\begin{Theorem}[\cite{frigro99}]\label{theo:10}
  Let $D$ be a \PTIME-decidable class of structures of bound\-ed local tree-width.
  Then $\MC(\FO)|_D$ is in $\FPT$.
\end{Theorem}

\noindent
For $\lambda:\mathbb N\rightarrow\mathbb N$ we let 
\[
\textit{GL}(\lambda):=\big\{\mathcal G\in\GRA\bigmid\forall \mathcal
H\preceq \mathcal G:\;\ltw(\mathcal H)\le\lambda\big\},
\]
and, for $\mu\in\mathbb N$,
\[
B(\lambda,\mu):=\big\{\mathcal A\in\STR\bigmid\exists X\subseteq
A\;(|X|\le\mu\wedge\mathcal G(\mathcal A\setminus X)\in 
\textit{GL}(\lambda))\big\}.
\]
Note that the clique-number of a graph in $\textit{GL}(\lambda)$ is
at most $\lambda(1)+1$, thus the clique-number of a graph in
$B(\lambda,\mu)$ is at most $\mu+\lambda(1)+1$. This implies that
$B(\lambda,\mu)\subseteq\STR[\mu+\lambda(1)+1]$.\footnote{Cf.\ Footnote
  $^{\ref{fuss}}$ on Page \pageref{fuss}.}

\begin{Lemma}\label{lem:7}
  Let $\lambda:\mathbb N\rightarrow\mathbb N$ and $\mu\in\mathbb N$. Then
  $\MC(\FO)|_{B(\lambda,\mu)}$ is in $\FPT$.
\end{Lemma}

\proof
The class $\textit{GL}(\lambda)$ of graphs is minor closed and hence
PTIME-decidable by Theorem \ref{theo:9}. This implies that
$B(\lambda,\mu)$ is PTIME-decidable.

Then for $\mu=0$ the statement follows from Theorem \ref{theo:10}.
The case $\mu>0$ can be reduced to the case $\mu=0$ as follows: For
every $\mathcal A\in\mathcal B(\lambda,\mu)$ and sentence
$\phi\in\FO$, we define a structure $\mathcal A^*\in B(\lambda,0)$ and a
sentence $\phi^*\in\FO$ in such a way that $\mathcal \mathcal A\models\phi$ if,
and only if, $\mathcal A^*\models\phi^*$, and the mappings $\mathcal
A\mapsto\mathcal A^*$ and $\phi\mapsto\phi^*$ are computable in
polynomial time.

So suppose we are given $\mathcal A\in\mathcal B(\lambda,\mu)$ and $\phi\in \FO$. For simplicity, we
assume that their vocabulary consists of a single binary relation
symbol $E$. 

Let $X\subseteq A$ such that $|X|\le\mu$ and $\mathcal A\setminus X\in
B(\lambda,0)$. Such an $X$ can be computed in polynomial time because
the class $B(\lambda,0)$ is PTIME-decidable. Say, $X=\{a_1,\ldots,a_\mu\}$.

Let $\tau:=\{E,P_1,\ldots,P_\mu,Q_1,\ldots,Q_\mu,R_1,\ldots,R_\mu\}$,
where the $P_i$, $Q_i$, and $R_i$ are unary. We let $A^*:=A$, 
$E^{\mathcal A^*}:=E^{\mathcal A}\cap(A\setminus X)^2$, 
and, for $1\le i\le \mu$
\begin{align*}
P_i^{\mathcal A^*}&:=\{a_i\},\\
Q_i^{\mathcal A^*}&:=\{b\in A\mid E^{\mathcal A}a_ib\},\\
R_i^{\mathcal A^*}&:=\{b\in A\mid E^{\mathcal A}ba_i\}.
\end{align*}
Furthermore, we let $\phi^*$ be the sentence obtained from $\phi$ by
replacing each subformula $Exy$ by
\[
Exy\vee\bigvee_{i=1}^\mu((P_ix\wedge Q_iy)\vee(P_iy\wedge R_ix)).
\]
Clearly, these definitions lead to the desired result.
\proofend

To complete the proof of Theorem \ref{theo:8} we use a decomposition
theorem for non-trivial minor-closed classes of graphs that roughly
says that all graphs in such a class are built up in a tree-like
manner from graphs in a $B(\lambda,\mu)$. It is based on Robertson and Seymour's deep
structure theory for graphs without $\mathcal K_n$-minors.
The precise statement
requires some new notation.
Let $(\mathcal T,(A_t)_{t\in T})$ be a tree-decomposition of a structure
$\mathcal A$. The \emph{torso} of this decomposition at $t\in\mathcal T$,
denoted by $[A_t]$, is the
graph with universe $A_t$ and an edge between two distinct
vertices $a,b\in A_t$ if either there is an edge between $a$ and $b$
in the Gaifman graph $\mathcal G(\mathcal A)$ or there exists an
$s\in T\setminus\{t\}$ such that $a,b\in A_s$. $(\mathcal
T,(A_t)_{t\in T})$ is a tree-decomposition \emph{over} a class $D$ of
graphs if all its torsos belong to $D$.

\begin{Theorem}[\cite{gro-b}]\label{theo:11}
  Let $D$ be a class of graphs with an excluded minor. Then there
  exist $\lambda:\mathbb N\rightarrow\mathbb N$ and $\mu\in\mathbb N$
  such that every $\mathcal G\in D$ has a tree-decomposition over
  $B(\lambda,\mu)$.
 
  Furthermore, given $\mathcal G$ such a decomposition can be computed in
  \PTIME.
\end{Theorem}

Clearly, this theorem implies the analogous statement for all
structures in $\STR[D]$.

The \emph{adhesion} of a tree-decomposition $(\mathcal T,(A_t)_{t\in
  T})$ is $\max\{|A_s\cap A_t|\mid E^{\mathcal
  T}st\}$. The \emph{clique number} of a class of structures is the
maximum of the clique numbers of the Gaifman graphs of structures in $D$,
if this maximum exists, or $\infty$ otherwise. Note that if
$(\mathcal T,(A_t)_{t\in T})$ is a decomposition over a class $D$ then
the clique-number of $D$ is an upper bound for the adhesion of
$(\mathcal T,(A_t)_{t\in T})$. Remembering that the clique-number of
$B(\lambda,\mu)$ is $\lambda(1)+\mu+1$, we see that the adhesion of a
tree-decomposition over $B(\lambda,\mu)$ is at most
$\lambda(1)+\mu+1$.
 
\medskip\noindent \textit{Proof (of Theorem \ref{theo:8}):}
Let $D$ be a \PTIME-decidable class of graphs with an excluded minor
and  $\lambda,\mu$ such that every $\mathcal G\in
D$ has a tree-decomposition over $B(\lambda,\mu)$. Let
$\nu:=\lambda(1)+\mu+1$.

We shall describe an algorithm that, given $\mathcal A\in \STR[D]$ and
$\phi\in\FO$, decides if $\mathcal \mathcal A\models\phi$.

So let $\mathcal A\in\STR[D]$, say, of vocabulary $\tau$ and
$\phi\in\FO[\tau]$.  Our algorithm starts by computing a
tree-decomposition $(\mathcal T,(A_t)_{t\in T})$ of $\mathcal A$ over
$B(\lambda,\mu)$.  For $t\in T$ we let $A_{\ge t}:=\bigcup_{u\ge
  t}A_u$ ($u\ge t$ if there is a path from $t$ to $u$ in the (directed)
tree $\mathcal T$).  In particular, $A_{\ge r}=A$ for the root
$r:=r^{\mathcal T}$ of $\mathcal T$.

Furthermore, we let $B_{r}:=\emptyset$ and $B_t:=A_t\cap A_s$ for
$t\in T\setminus\{r\}$ with parent $s$. Recall that the adhesion of
$(\mathcal T,(A_t)_{t\in T})$ is at most $\nu$. Thus $|B_t|\le\nu$ for
$t\in T$.

The \emph{quantifier rank} of a first-order formula is the maximal
depth of nested quantifiers in this formula. Let $q$ be the quantifier
rank of $\phi$. Simple techniques from logic show that there is an
algorithm that, given a vocabulary $\tau$ and $q,m\in \NN$, computes a
finite set $\Phi_{\tau,q,m}$ of first-order formulas of vocabulary
$\tau$ of quantifier rank $\le q$ with free variables among
$v_1,\ldots,v_m$ such that every such formula is equivalent to a
formula in $\Phi_{\tau,q,m}$. Without loss of generality we can assume
that $\phi\in\Phi_{\tau,q,m}$ (otherwise we can compute a
$\phi'\in\Phi_{\tau,q,m}$ equivalent to $\phi$ and work with $\phi'$).

A $(\tau,q,m)$-\emph{type} is a subset of $\Phi_{\tau,q,m}$. Given a
$\tau$-structure ${\mathcal A}'$ and $B=\{b_1,\ldots,b_m\}\subseteq
A'$ let $\text{tp}^{\mathcal A'}_q(B)$ -- more precisely,
$\text{tp}^{\mathcal A'}_q(b_1,\ldots,b_m)$ -- be the
$(\tau,q,m)$-type
\[
\text{tp}^{\mathcal A'}_q(B):=\{\psi(v_1,\ldots,v_m) \in \Phi_{\tau,q,m} \mid {\mathcal A}' \models \psi(b_1,\ldots,b_m)\}.
\]
We come back to our structure ${\mathcal A}$ and the tree-decomposition
$(\mathcal T,(A_t)_{t\in T})$.  By induction from the leaves to the
root, for every $t\in T$ we compute $\text{tp}^{\langle A_{\ge
    t}\rangle^{\mathcal A}}_q(B_t)$, which for brevity we denote by
$\text{tp}^{\ge t}_q(B_t)$.  Since $B_r=\emptyset$,
$\text{tp}^{\ge r}_q(B_r)$ is a set of sentences, and we have
$\mathcal A\models\phi$ if, and only if, $\phi\in\text{tp}^{\ge r}_q(B_r)$.

\medskip
So let $t$ be a vertex of $\mathcal T$ and assume that we have
already computed $\text{tp}^{\ge u}_q(B_u)$ for all children $u$ of $t$ (if
there are any). (Actually, the case that $t$ has no children is much
simpler than the following general case, because it is a direct
application of  Lemma \ref{lem:7}.)

For every $(\tau,q,m)$-type $\Phi$ we introduce a new $(m+1)$-ary
relation symbol $R_{\Phi}$. Furthermore, we let $P_1,\ldots,P_\nu$ be
new unary relation symbols and $\tau':=\tau\cup \{R_{\Phi} \mid m \le
\nu,\, \Phi \mbox{ a } (\tau,q,m)\mbox{-type}\}\cup\{P_1,\ldots,P_\nu\}$.
In four steps, we define a $\tau'$-structure $\widetilde{\mathcal A}_t$
that contains all the relevant information to compute $\text{tp}^{\ge
  t}_q(B_t)$. In the first three steps we define ``intermediate''
structures $\mathcal A_t^1$, $\mathcal A_t^2$, $\mathcal A_t^3$.
\begin{enumerate}
\item
$\mathcal A_t^1$ is the induced substructure of $\mathcal A$ with universe
$A_t$.
\item Suppose that $B_t=\{b_1,\ldots,b_m\}$ for an $m\le\nu$. Then
  $\mathcal A_t^2$ is the $\tau \cup \{P_1,$ $ \ldots,$ $ P_\nu\}$-expansion of
  $\mathcal A_t^1$ with $P_i^{\mathcal A_t^2}:=\{b_i\}$ for $1\le i\le
  m$ and $P_i^{\mathcal A_t^2}:=\emptyset$ for $m+1\le i\le \nu$. 
\item
${\mathcal A}_t^3$ is obtained from ${\mathcal
  A}_t^2$ by adding a new vertex $c_u$ for
  every child $u$ of $t$ and edges from $c_u$ to all vertices of
  $B_u$.
\item
$\widetilde{\mathcal A}_t$ is the $\tau'$-expansion of
$\widetilde{\mathcal A}_t^3$ with
\begin{align*}
R_\Phi^{\widetilde{\mathcal
 A}_t}:=\big\{(d_1,\ldots,d_{m_u},c_u)\bigmid&
u\text{ child of }t,\\
&B_u=\{d_1,\ldots,d_{m_u}\}, \text{tp}^{\ge u}_q(B_u)=\Phi\big\}.
\end{align*}
\end{enumerate}
Standard Ehrenfeucht-Fra\"{\i}ss\'e type methods show that there is a
computable function that associates with every formula
$\psi\in\Phi_{\tau,q,m}$ a sentence
$\widetilde{\psi}\in\FO[\tau']$ such that
$\psi\in\text{tp}^{\ge t}_q(B_t)$ if, and only if,
$\widetilde{\mathcal A}_t\models\widetilde{\psi}$.

We claim that $\widetilde{\mathcal A}_t\in B(\lambda+1,\mu)$. To
see this, observe that the Gaifman graph $\mathcal G({\widetilde{\mathcal
    A}_t})$ is the graph obtained from the torso
$[A_t]$ by adding the vertices $c_u$ and edges between $c_u$ and every
element of $B_u$. Recall that, by the definition of the torso, each $B_u$ is a
clique in $[A_t]$. It is easy to see that adding
vertices and connecting them with cliques can increase the
tree-width of a graph by at most one. This implies the claim.

Now we can put everything together and  obtain an algorithm deciding $\MC(\FO)_{\STR[D]}$, a high-level description of which is
given as Algorithm \ref{alg:2}.
\begin{figure}
\begin{center}
  \fbox{\parbox{11cm}{ \setcounter{zeilencounter}{0}
      \textsc{ModelCheck$_D$}($\mathcal A\in \STR$, $\phi\in\FO$)
\begin{tabbing}
\hspace*{4mm}\=\hspace{9mm}\=\hspace{6mm}\=\hspace{6mm}\=\hspace{6mm}\=\hspace{6mm}\=\hspace{6mm}\=\hspace{6mm}\=\kill
\zeile \IF\ $\mathcal G(\mathcal A)\not\in D$ \THEN\ \REJECT\\
\zeile compute tree-decomposition $(\mathcal T,(A_t)_{t\in T})$ of
$\mathcal A$ over $B(\lambda,\mu)$\\
\zeile $q:=\text{qr}(\phi)$, $\tau:=\text{vocabulary of }\phi$\\
\zeile \FOR\ $m=0$ \TO\ $\nu$\\
\zeile \>compute $\Phi_{\tau,q,m}$\\
\zeile \FORALL\ $t\in T$ (from the leaves to the root)\\
\zeile \>compute $\widetilde{\mathcal A}_t$\\
\zeile \>$\text{tp}^{\ge t}_q(B_t):=\emptyset$\\
\zeile \>$m:=|B_t|$\\
\zeile \>\FORALL\ $\psi\in\Phi_{\tau,q,m}$\\
\zeile \>\>compute $\widetilde{\psi}$\\
\zeile \>\>\IF\ $\widetilde{\mathcal A}_t\models\widetilde{\psi}$\\
\zeile \>\>\>\THEN\ $\text{tp}^{\ge t}_q(B_t):=\text{tp}^{\ge
  t}_q(B_t)\cup\{\psi\}$\\
\zeile \IF\ $\phi\in\text{tp}^{\ge r}_q(B_r)$\\
\zeile \>\THEN\ \ACCEPT\\
\zeile \>\ELSE\ \REJECT.
\end{tabbing}}}
\end{center}
\alg\label{alg:2}
\end{figure}
Its correctness is straightforward. Let us just have a look at the
running time: Let $n$ be the size of the input structure. Then Lines 1
and 2 require time polynomial in $n$ (indendently of $\phi$). The time
required in Lines 3--5 only depends on $\phi$. The main loop in Lines
6--13 is called $|T|$ times, which is polynomial in $n$. Computing
$\widetilde{\mathcal A}_t$ is polynomial in $|A_t|$ and the number of
children of $t$ since we have already computed $\text{tp}^{\ge
  u}_q(B_u)$ for all children $u$ of $t$ (with constants heavily
depending on $||\phi||$). The main task is to decide whether
$\widetilde{\mathcal A}_t\models\widetilde{\psi}$ in Line 12; by Lemma
\ref{lem:7} this is fixed-parameter tractable because
$\widetilde{\mathcal A}_t\in B(\lambda+1,\mu)$. The time required in
Lines 14--16 again only depends on $\phi$.  \proofend

A consequence of our results is that slicewise first-order definable
parameterized problems are fixed-parameter tractable when restricted to
classes of structures whose underlying class of graphs has an excluded minor.

\begin{Corollary}\label{cor:6}
  Let $D$ be a \PTIME-decidable class of graphs with an excluded minor and
  $P\subseteq\STR\times\Pi^*$ a parameterized problem that is
  slicewise \FO-definable. Then $P|_{\STR[D]}$ is in \FPT.
\end{Corollary}

\section{A logical characterization of fixed-parameter tractability}\label{sec:8}
In this section we give a characterization of \FPT\ in the spirit of
descriptive complexity theory.

We briefly review some facts from this area (see
\cite{ebbflu95,imm99} for details). It
is common in descriptive complexity theory to identify decision
problems, usually modeled
by languages $L\subseteq\Sigma^*$ for a finite alphabet $\Sigma$, with
classes of finite structures. More precisely, one identifies problems
with classes of \emph{ordered finite structures}. An \emph{ordered
  structure} is a structure whose vocabulary contains the binary
relation symbol $\le$, and this symbol is interpreted as a linear order
of the universe. \ORD\ denotes the class of all ordered structures. In
this section, $\tau$ always denotes a vocabulary that contains $\le$.

One of the most important results in descriptive complexity theory is
the Immerman-Vardi Theorem \cite{imm86,var82} saying that a class of
ordered structures is in PTIME if, and only if, it is definable in
least-fixed point logic FO(LFP). More concisely,
\[
\PTIME=\FO(\LFP).
\]
We prove a similar result characterizing the class \FPT\ in terms of
the finite variable least fixed-point logics $\LFP^s$, for $s\ge 1$,
which were introduced by Kolaitis and Vardi \cite{kolvar96}. Analogously to
the classical setting we model parameterized problems by subsets
of $\ORD[\tau]\times\mathbb N$, for some $\tau$.

\begin{Theorem}\label{theo:12}
  A parameterized problem $P\subseteq\ORD[\tau]\times\mathbb N$ is in \FPT\
  if, and only if, there is an $s\ge 1$ such that $P$ is slicewise
  $\LFP^s$-definable. More concisely, we may write
\[
\FPT=\bigcup_{s\ge 1}{\rm slicewise\mbox{-}}\LFP^s.
\]
\end{Theorem}

In the proof of this result we assume that the reader is familiar with
descriptive complexity theory, in particular with least fixed-point
logic and the proof of the Immerman-Vardi Theorem. Those who are not
may safely skip the rest of this section.

We first recall the definition of $\LFP^s$: In the terminology of
\cite{ebbflu95} (p.\ 174), $\LFP^s$-sentences are \FO(\LFP)-sentences in the form
\[
\exists\bar y[\text{S--LFP}_{\bar x_1,X_1,\ldots,\bar
  x_m,X_m}\phi_1,\ldots,\phi_m]\bar y,
\]
where $\phi_1,\ldots,\phi_m$ are first-order formulas with at most $s$
individual variables. (That is, $\LFP^s$-sentences are existential closures of
simultaneous fixed-points over $\FO^s$-formulas.)

We use the following two facts, the first implicit in \cite{var95}
and the second in the proof of the Immerman-Vardi
Theorem (cf.\ \cite{ebbflu95}). Fix $\tau$ and $s\in\mathbb N$ and let
$n$ always denote the size of the input structure.
\begin{enumerate}
\item There is a computable function that associates an
  $O(n^{2s})$-algorithm $\mathbb A_\phi$ with each $\phi\in\LFP^s[\tau]$
  such that $\mathbb A_\phi$ accepts a structure $\mathcal
  A\in\ORD[\tau]$ if, and only if, $\mathcal A$ satisfies $\phi$.
\item There is a $t\in\mathbb N$ and a
  computable function that associates with every $O(n^s)$-algorithm
  $\mathbb A$ accepting a class $C\subseteq\ORD[\tau]$ a sentence
  $\phi_{\mathbb A}\in\LFP^t[\tau]$ such that a $\tau$-structure
  $\mathcal A$ satisfies $\phi_{\mathbb A}$ if, and only if, $\mathcal
  A\in C$.
\end{enumerate}

There is a slight twist in (2). When proving it one usually
assumes that all structures are sufficiently large, in particular
larger than the constant hidden in $O(n^s)$. This way it can be
assumed that the algorithm is actually an $n^{s+1}$-algorithm (without
any hidden constants.) Then one argues that small structures are no
problem because they can be described up to isomorphism in first-order
logic. When restricting the number of variables, one has to be careful
with such an argument. Luckily, we are safe here
because we only consider ordered structures, and there is a
$t\in\mathbb N$ (depending on $\tau$) such that every structure $\mathcal
A\in\ORD[\tau]$ can be characterized up to isomorphism by an $\FO^t$-sentence.

\medskip\noindent
\textit{Proof (of Theorem \ref{theo:12}):}
For the backward direction, suppose that $P\subseteq\ORD[\tau]\times\mathbb
N$ is slicewise $\LFP^s$-definable via $\delta:\mathbb
N\rightarrow\LFP^s$. Then Algorithm \ref{alg:3} shows that $P$ is in \FPT. The crucial fact is that Lines
1 and 2 do not depend on the input structure $\mathcal A$ and Line 3
requires time $O(n^{2s})$.

\begin{figure}
\begin{center}
  \fbox{\parbox{11cm}{ \setcounter{zeilencounter}{0}
      \textsc{Decide-P}($\mathcal A\in\ORD$, $k\in\mathbb N$)
\begin{tabbing}
\hspace*{4mm}\=\hspace{9mm}\=\hspace{6mm}\=\hspace{6mm}\=\hspace{6mm}\=\hspace{6mm}\=\hspace{6mm}\=\hspace{6mm}\=\kill
\zeile compute $\delta(k)\in\LFP^s$\\
\zeile compute $\mathbb A_{\delta(k)}$ (cf. (1))\\
\zeile simulate $\mathbb A_{\delta(k)}$ on input $\mathcal A$\\
\zeile \IF\ $\mathbb A_{\delta(k)}$ accepts $\mathcal A$\\
\zeile\> \THEN\ \ACCEPT\\
\zeile\> \ELSE\ \REJECT.
\end{tabbing}}}
\end{center}
\alg\label{alg:3}
\end{figure}

For the forward direction, suppose that $P\subseteq\ORD[\tau]\times\mathbb
N$ is in \FPT. Choose $f:\mathbb N\rightarrow\mathbb N$, $c\in\mathbb
N$ and an algorithm $\mathbb A$ deciding $P$ in time
$f(k)\cdot n^c$. The algorithm $\mathbb A$ gives rise to a sequence
$\mathbb A_k$ ($k\ge 1$) of algorithms, where $\mathbb A_k$ decides
the class $\{\mathcal A\mid(\mathcal A,k)\in P\}\subseteq\ORD[\tau]$
in time $O(n^c)$. Then (2) yields the desired slicewise definition of
$P$.
\proofend

\section{Beyond FPT}\label{sec:9}
In this last section we discuss how logical definability is related to the
classes of the W-hierarchy. We introduce another hierarchy of parameterized
problems, which we call the A-hierarchy, in terms of model-checking problems
for first-order logic and show that the A-hierarchy can be seen as a
parametric analogue of the polynomial hierarchy. We then discuss the relation
between the A-hierarchy and the W-hierarchy.

Our treatment is motivated by the following two results. They relate the class
$\W[1]$ to model-checking and computations of non-deterministic Turing
machines, respectively. Recall that $\MC(\Sigma_1[s])$ denotes the
parameterized model-checking problem for existential formulas in prenex normal
form whose vocabulary contains at most $s$-ary relation symbols.

\begin{Theorem}[Downey, Fellows, Regan \cite{dowfelreg98}]\label{theo:4}
$\MC(\Sigma_{1})|_{\GRA}$ is $\W[1]$-complete under parameterized m-reductions.
\end{Theorem}

\proof
We prove that $\CLI\eprm\MC(\Sigma_{1})|_{\GRA}$.

$\CLI\prm\MC(\Sigma_{1})|_{\GRA}$ follows from the fact that $\CLI$ is
slicewise $\Sigma_1$-definable, so we only have to prove the
converse. 

An \emph{atomic $k$-type} (in the theory of graphs) is a sentence
$\theta(x_1,\ldots,x_k)$ of the form $\bigwedge_{1\le i<j\le k}\alpha_{ij}(x_i,x_j)$,
where $\alpha_{ij}(x_i,x_j)$ is either $x_i=x_j$ or $Ex_ix_j$ or
$(\neg Ex_ix_j\wedge \neg x_i= x_j)$ (for $1\le i< j\le k$).

It is easy to see that there is a computable mapping $f$ that
associates with every \EFO-sentence $\phi$ a sentence $\tilde{\phi}$ of the
form
\begin{equation}\label{eq2.1}
\bigvee_{i=1}^l\exists x_1\ldots \exists x_k\theta_i(x_1,\ldots,x_k),
\end{equation}
where each $\theta_i$ is an atomic $k$-type, such that for all graphs
$\str G$ we have $\str G\models\phi\iff \str G\models\tilde{\phi}$.

For each graph $\str G$ and each atomic $k$-type $\theta(\bar
x)=\bigwedge_{1\le i<j\le k}\alpha_{ij}(x_i,x_j)$ we define a graph
$h(\str G,\theta)$ as follows:
\begin{itemize}
\item
The universe of $h(\str G,\theta)$ is $\{1,\ldots,k\}\times G$.
\item
There is an edge between $(i,v)$ and $(j,w)$, for $1\le i<j\le k$ and
$v,w\in G$, if $\str G\models\alpha_{ij}(v,w)$.
\end{itemize}
Then $h(\str G,\theta)$ contains a $k$-clique if, and only if,
$\str G\models\exists\bar x\theta(\bar x)$.
Now we are ready to define the reduction from
$\MC(\Sigma_1)|_{\GRA}$ to $\CLI$. 
Given an instance $(\mathcal G,\phi)$ of
$\MC(\Sigma_1)|_{\GRA}$, we first compute the sentence
\[
\tilde{\phi}=\bigvee_{i=1}^l\exists x_1\ldots\exists
x_k\theta_i.
\] 
We let $\mathcal G'$ be the disjoint union of the
graphs $h(\mathcal G,\theta_i)$ for $1\le i\le l$. Then $\mathcal G'$
has a $k$-clique if, and only if, $\mathcal G\models\phi$.
\proofend

\begin{Theorem}[Cai, Chen, Downey, and Fellows \cite{caichedowfel97}]\label{theo:13}
  The  parameterized problem \textit{SHORT
  TURING MACHINE AC\-CEPT\-ANCE (NM)\footnote{\textit{NM}\/ stands for
  \underline{n}ondeterministic Turing \underline{m}achine. This notation
  should be seen in connection with the $\AM_t$ below, which refers to \underline{a}lternating Turing \underline{m}achines.}} is $\W[1]$-complete, where

\PP{NM}{A non-deterministic Turing machine $M$}{$k\in\mathbb N$}{Does
  $M$ accept the empty word in at most $k$ steps}
\end{Theorem}

Downey and Fellows call Theorem \ref{theo:13} a parameterized ``analog of
Cook's Theorem''.  In our notation, we may write $\mclass{\NM}=\W[1]$.  It is
now very natural to define a ``parameterized analogue of the polynomial
hierarchy'', which we call the \emph{$\A$-hierarchy}, by letting
$\A[t]:=\mclass{\AM_t}$ for all $t\ge 1$, where

\PP{$\AM_t$}{An alternating Turing machine $M$ whose initial
  state is existential}{$k\in\mathbb N$}{Does $M$
  accept the empty word in at most $k$ steps with at most $t$ alternations}

\noindent
Note that $\NM=\AM_1$, thus $\W[1]=\A[1]$.
Our following theorem can be seen as a natural generalization of
Theorem \ref{theo:4}. 

\begin{Theorem}\label{theo:14}
  For all $t\ge 1$, the problem $\MC(\Sigma_t)|_{\GRA}$ is
  $\A[t]$-complete under parameterized m-reductions. Thus
\[
\A[t]=\mclass{\MC(\Sigma_t)|_{\GRA}}=\bigcup_{s\ge 1}\mclass{\MC(\Sigma_t[s])}.
\]
\end{Theorem}

\proof
The second equality follows from Corollary \ref{cor:1}(3).

To prove that $\MC(\Sigma_t)|_{\GRA}\prm\AM_t$, we first observe that,
for every graph $\mathcal G$ and every quantifier-free formula
$\theta(x_1,\ldots,x_m)$, 
in time 
$p(||\theta||)\cdot|G|^2$, for
 a suitable
polynomial $p$, we can construct
 a deterministic Turing machine 
$M({\mathcal G},\theta)$ with input alphabet $G$ that accepts an input
word $a_1\ldots a_m$ over  $G$ if, and only if, $\mathcal
G\models\theta(a_1,\ldots,a_m)$
and that performs at most $f(||\theta||)$  steps for some computable function
 $f:\mathbb N\rightarrow\mathbb N$.
Just to give an example, to check
whether $Ex_kx_l$ holds, say with $k<l$, we need states $s(i)$ for
$1\le i\le k$ and $s(i,a)$ for $k< i\le l, a\in G$. The machine
starts  in state $s(1)$ with head in position 1 in state $s(1)$. It moves 
its head right until it reaches
position $k$ in state $s(k)$, reads $a_k$ and goes to position $(k+1)$
in  state $s(k+1,a_k)$. Then it moves right again until it reaches
position $l$ in state $s(l,a_k)$. From this state it can reach an
accepting state if, and only if, $E^{\mathcal G}a_ka_l$.

Now suppose we are given an instance of $\MC(\Sigma_t)|_{\GRA}$, i.e.\
a graph $\mathcal G$ and a sentence
\[
\phi=\exists x_{11}\ldots\exists x_{1k_1}\forall x_{21}\ldots\forall
x_{2k_2}\;\ldots\; Qx_{t1}\ldots Qx_{tk_t}\;\theta,
\]
where $\theta$ is quantifier-free. Then the following 
alternating Turing machine $M(\mathcal G,\phi)$  accepts the empty word
if, and only if, $\mathcal G\models\phi$: It first writes a sequence
of elements of $\mathcal G$ on the tape using existential and
universal states appropriately and then simulates $M(\mathcal
G,\theta)$ on this input. Again $g(||\theta||)$, for some computable function $g$, is an upper bound
for the number of steps $M(\mathcal G,\phi)$ has to perform.

To finish the proof, by Corollary \ref{cor:1}(3) it suffices to show that
$\text{AM}_t\prm\MC(\Sigma_t[\tau])$ for a suitable vocabulary $\tau$.
To illustrate the idea, we first consider the case $t=1$.  Suppose
we are given a nondeterministic Turing machine $M$ with alphabet $\Sigma$,
set $Q$ of states, initial state $q_{0}$, accepting state
$q_{\textup{acc}}$ and transition relation $\delta$. Let $\Sigma_{H}:=
\{a_{H} \mid a \in \Sigma\}$, $a_{H}$ coding the information that the
head of $M$ scans a cell containing $a$. Let $\tau:= \{\textit{ST},
\textit{AL}, H, \textit{IN}, \textit{ACC}, R,L,S\}$ with unary
$\textit{ST}, \textit{AL}, H, \textit{IN}, \textit{ACC}$ and 4-ary
$R,L,S$ and let ${\cal{A}}_{M}$ be the $\tau$-structure given by
$$
\begin{array}{lll}
\multicolumn{3}{l}{A_M:=Q \dot{\cup}\Sigma \dot{\cup} \Sigma_H;}\\
\textit{ST}^{\cal{A}_M}:= Q;&\textit{AL}^{\cal{A}_M}:= \Sigma;&H^{\cal{A}_M}:=\Sigma_H;\\
\textit{IN}^{\cal{A}_M}:=\{q_0\};&\textit{ACC}^{\cal{A}_M}:=\{q_{\textup{acc}}\};&\\
\multicolumn{3}{l}{R^{\cal{A}_M}:=\{(q,a_H,b,q') \mid (q,a,1,b,q')\in \delta \};}\\
\multicolumn{3}{l}{L^{\cal{A}_M}:=\{(q,a_H,b,q') \mid (q,a,-1,b,q')\in \delta \};}\\
\multicolumn{3}{l}{S^{\cal{A}_M}:=\{(q,a_H,b_H,q') \mid (q,a,0,b,q')\in \delta \}
\cup \{(q_{\textup{acc}},a_H,a_H,q_{\textup{acc}}) \mid a\in \Sigma\},}
\end{array}
$$
where $(q,a,h,b,q') \in \delta$ means: if $M$ is in state $q$ and its head
scans $a \in \Sigma$, then $M$ replaces $a$ by $b$, moves its head one cell
 to
the right $(h = 1)$, to the left $(h = -1)$, or does not move its head 
$(h =
0)$; finally, it changes to state $q'$.

Let $k$ be given as parameter for $\textit{AM}_{1}$. In $k$ steps, $M$ scans at
most the first $k$ cells. The quantifier-free formula (note that the following
formulas only depend on $k$ and not on $M$)
\[
\phi_{\textup{config}}
(x,y_{1}, \ldots, y_{k}) := \textit{ST}x \wedge \bigvee_{i=1}^k(Hy_i \wedge 
\bigwedge_{j\neq i}\textit{AL}y_j)
\]
states that $(x,y_{1}, \ldots, y_{k})$ is a configuration with state
$x$, with the head facing $y_{i}$, and with $y_j$ being the content of
the $j$th cell $(j\not= i)$.  Let $\phi_{\textup{start}}(x,y_{1},
\ldots, y_{k})$ be a quantifier-free formula stating that $(x,y_{1},
\ldots, y_{k})$ is the starting configuration.  Similarly, we define a
quantifier-free formula $\phi_{\textup{step}} (x,y_{1}, \ldots,
y_{k},x',y'_{1}, \ldots, y'_{k})$ stating that the configuration
$(x',y'_{1}, \ldots, y'_{k})$ is the successor configuration of
$(x,y_{1}, \ldots , y_{k})$ (we agree that each accepting
configuration is its own successor).

 Now, the equivalence
\begin{eqnarray*}
\mbox{$M$ stops in $\le k$ steps}&\iff&{\mathcal A}_M\models \phi _k
\end{eqnarray*}
holds for the existential sentence $\phi _k$:
$$
\begin{array}{l} \exists x_1\exists y_{11}\ldots \exists y_{1k}\;  \ldots \; 
\exists x_k\exists y_{k1}\ldots\exists y_{kk}(\phi_{\textup{start}} (x,y_{11}, \ldots, 
y_{1k})\\
\ \  \wedge \bigwedge_{i=1}^{k-1}
\phi_{\textup{step}} (x_i,y_{i1}, \ldots, y_{ik},x_{i+1},y_{i+11}, \ldots,
y_{i+1k}) \wedge \textit{ACC}x_k).
\end{array}
$$
For $t\ge 1$, we can proceed similarly, with the addition that
universal quantifiers are needed to take care of universal states of
the input machine. Now, $\tau$ also contains 
two further unary
symbols $F$ (for the existential states) and $U$ (for the universal states)
which get the corresponding interpretations in ${\cal{A}}_{M}$. For
example, for $t=2$, we can take a $\Sigma_{t}$-sentence7
equivalent to:
{\renewcommand{\arraystretch}{1.3}
\[
\begin{array}{l} 
\displaystyle\bigvee_{l\le k} 
\exists x_1\, \exists y_{11}\ldots \exists y_{1k}\,\ldots\, 
\exists x_k\exists y_{l1}\ldots\exists y_{lk}\Big(\\
\begin{array}{c@{\,}l@{}}
&\phi_{\textup{start}} (x,y_{11},\ldots,y_{1k})\\
\wedge&
\bigwedge_{i=1}^{l-1}\phi_{\textup{step}}(x_i,y_{i1},\ldots,y_{ik},x_{i+1},y_{i+1\, 1},\ldots,y_{i+1\, k})\\
\wedge&
Fx_1 \wedge \ldots \wedge Fx_{l-1}\wedge Ux_l\vspace{1mm}\\
\wedge&
\forall x_{l+1}\forall y_{l+1\, 1}\ldots\forall y_{l+1\, k}\,\ldots\,
\forall x_k\, \forall  y_{k1}\ldots \forall  y_{kk}\big(\\
&\hspace{5mm}\begin{array}[t]{l@{\,}l@{}}
\big(&Ux_{l+1}\wedge \ldots \wedge Ux_{k-1}\\
\wedge&\bigwedge_{i=l}^{k-1}\phi_{\textup{step}}(x_i,y_{i1},\ldots,y_{ik},x_{i+1},y_{i+1\, 1}, \ldots,
y_{i+1\, k}) \big) \rightarrow
\textit{ACC}x_k)\big)\Big).
\end{array}
\end{array}
\end{array}
\]}
(Without loss of generality we assume that the accepting state is
both existential and universal and that at least one transition is
always possible in a universal state). 
\proofend

The following result due to Downey, Fellows, and Regan allows to
compare the $\W$- and the $\A$-hierarchy. Let $t, u \ge 1$. A formula
$\phi$ is $\Sigma_{t,u}$, if it is $\Sigma_{t}$ and all quantifier
blocks after the leading existential block have length $\le u$.

\begin{Theorem}[\cite{dowfelreg98}]\label{theo:15}
For all $t\ge 1$,
\[
\W[t]=\bigcup_{\substack{\tau\textup{ vocabulary}\\u\ge 1}}\mclass{\MC(\Sigma_{t,u}[\tau])}=\bigcup_{u\ge 1}\mclass{\MC(\Sigma_{t,u})|_{\GRA}}.
\]
\end{Theorem}

Note that for $t=1$, this is just Theorem \ref{theo:4}. The crucial step in
proving the theorem for $t\ge 2$ is to establish the $\W[t]$-completeness of
the problems \textit{WEIGHTED MONOTONE $t$-NORMALIZED SATISFIABILITY} (for
even $t$) and \textit{WEIGHTED ANTIMONOTONE $t$-NORMALIZED SATISFIABILITY}
(for odd $t\ge 3$). We refer the reader to \cite{dowfel99} for the (difficult)
proofs of these results. Once these basic completeness results are
established, it is relatively easy to derive Theorem \ref{theo:15} (also cf.\ 
our proof of Theorem \ref{theo:17}). We encourage the reader to give a purely
``logical'' proof of the second equality in the theorem.

Since $\Sigma_{1,u}=\Sigma_{1}$, we get 
\[
\W[1]=\bigcup_{\tau\text{ vocabulary}}\mclass{\MC(\Sigma_{1}[\tau])}=
\mclass{\MC(\Sigma_{1})|_{\GRA}}=\A[1].
\]
By Theorem \ref{theo:14},  $\W[t] \subseteq \A[t]$ for all
$t\ge 2$.
The question whether $\W[t] = \A[t]$ for all $t$ 
remains open; in view of Theorem \ref{theo:14} this question is equivalent to
$\W[t]=\mclass{\MC(\Sigma_{t})|_{\GRA}}$ for all $t\ge 1$. In this form it is stated as an open
problem in \cite{dowfelreg98}.
 Consider, for example, the following parameterized problem

\PP{P$_0$}{Graph $\mathcal G$}{$(k,l)\in\mathbb N\, ^2$}{Are there
  $a_1,\ldots,a_k\in G$ such that every clique of size $l$ contains
  an $a_i$}

\noindent
Since $P_0$ is slicewise $\Sigma_2$-definable, we have $P_0\in\A[2]$.
But is $P_0$ in $\W[2]$?

In the 
definition of the $\W$-hierarchy we can restrict the length of the 
non-leading quantifier-blocks to one:

\begin{Proposition}\label{prop:1}
For all $t\ge 1$,
\[
\W[t]=\bigcup_{\tau\textup{ vocabulary}}\mclass{\MC(\Sigma_{t,1}[\tau])}.
\]
\end{Proposition}

\proof
The inclusion $\supseteq$ being trivial we turn to a proof of
$\subseteq$: Fix $\tau$ and $u \ge 1$. We show that
$\MC(\Sigma_{t,u}[\tau])\prm\MC(\Sigma_{t,1}[\tau'])$
for suitable $\tau'$.  The idea is to replace the blocks of at most $u$ 
quantifiers by a single quantifier
ranging over the set of $u$-tuples of a structure.  
 To explain this idea we first use a vocabulary  
containing function symbols (and sketch afterwards how one can do 
without).
Let $\tau': = \tau \cup
\{T, p_{1},\ldots,p_{u}\}$, where $T$ is a unary relation symbol (for ordered
$u$-tuples) and $p_{1},\ldots,p_{u}$ are unary function symbols (the
projection functions). Given a $\tau$-structure ${\cal{A}}$ let 
${\cal{A}}'$ 
be a $\tau'$-structure with
\begin{eqnarray*}
A':=A\dot{\cup}A^u,& \ \ &T^{\mathcal A'}:=A^u,
\end{eqnarray*}
where for $(a_1,\ldots,a_u)\in A^u$, $p_i(a_1,\ldots,a_u)=a_i$ and where the
relation symbols of $\tau$ are interpreted as in $\mathcal A$. Now, e.g.\ for
\[
\phi =\exists x_1\ldots \exists x_k\forall y_1\ldots \forall y_u
\psi(\bar{x},\bar{y})
\]
with quantifier-free $\psi$, let
$$
\phi'= \exists x_1\ldots \exists x_k\forall y(\neg Tx_1\wedge\ldots \wedge
\neg Tx_k \wedge (Ty \rightarrow \psi(\bar{x},p_1(y)\ldots p_u(y))).
$$
Then,
\[
{\mathcal A}\models \phi\iff{\mathcal A'}\models \phi',
\]
which gives the desired parameterized m-reduction.

Let us explain, for the case $t=2$,  how to proceed to avoid function symbols. One has to add to 
$\tau$, besides $T$ as 
above, for every relation symbol $R\in \tau$, say $r$-ary, every subset $M 
\subseteq \{1,\ldots,r\}$, and every function $\rho:M\rightarrow \{1,\ldots,u\}$ a 
new relation symbol
$R_{M,\rho}$; e.g., if $M=\{s,\ldots,r\}$ then
$R_{M,\rho}^{\mathcal A'}a_1\ldots a_{s-1}b$ if, and only if, $a_1\ldots a_{s-1}\in A$, 
$b=(b_1,\ldots,b_u)\in A^u$,  
and $R^{\mathcal A}a_1\ldots a_{s-1}b_{\rho(s)}\ldots b_{\rho(r)}$; and a subformula 
$Rx_{i_1}\ldots x_{i_{s-1}}y_{\rho(s)}\ldots y_{\rho(r)}$ of $\varphi $ is replaced by
$R_{M,\rho}x_{i_1}\ldots x_{i_{s-1}}y$.
\proofend

Downey, Fellows, and Regan \cite{dowfelreg98} also gave a (much simpler)
characterization of the $\W$-hierarchy in terms of Fagin-definability. We find
it worthwhile to sketch a short proof of this result. For a class $\Phi$ of
formulas we let $\textup{FD}(\Phi)$ be the class of all problems that are
$\Phi$-Fagin-definable. Let $\Pi_t[s]$ denote the class of all
$\Pi_t$-formulas whose vocabulary is at most $s$-ary.

\begin{Theorem}[\cite{dowfelreg98}]\label{theo:16}
  For all $t\ge 1$ we have
  $\W[t]=\mclass{\textup{FD}(\Pi_t[2])}=\mclass{\textup{FD}(\Pi_t)}$.
\end{Theorem}

\proof
Recall that $\W[t]=\bigcup_{d\ge1}\mclass{\textit{WSAT}(C_{t,d})}$, where
$C_{t,d}$ is the class of all propositional formulas of the form
\begin{equation}\label{eq:fdw}
\bigwedge_{i_1}\bigvee_{i_2}\ldots\underset{i_t}{(\bigwedge/\bigvee)}\;\phi_{i_1\ldots
  i_t}
\end{equation}
where the $\phi_{i_1\ldots i_t}$ are small formulas of depth at most $d$.

To prove that $\W[t]\subseteq\mclass{\textup{FD}(\Pi_t[2])}$, we first
transform a propositional formula $\phi$ of the form \eqref{eq:fdw}
into a propositional formula $\phi'$ of essentially the same form, but
with all the $\phi_{i_1\ldots i_t}$ being disjunctions (if $t$ is
even) or conjunctions (if $t$ is odd) of exactly $d'$ literals, for
some constant $d'$ only depending on $d$. This can be done by first
transforming the small formulas into equivalent formulas in
conjunctive normal form or disjunctive normal form, respectively, and
then repeatedly replacing disjunctions (conjunctions, respectively)
$\gamma$ with less than the maximum number of literals by the two clauses
$\gamma\vee X$ and $\gamma\vee\neg X$ ($\gamma\wedge X$ and
$\gamma\wedge\neg X$, respectively), for some variable $X$ not
appearing in $\gamma$.

We associate with $\phi'$ an $\{E,P,N,T,L\}$-structure $\mathcal C$ which is
obtained from the tree corresponding to $\phi'$ as follows: We first remove
the root. Then we identify all leaves corresponding to the same
propositional variable. To indicate whether a variable occurs positively or
negatively in a clause, we use the binary relations $P$ and $N$. The unary
relation $T$ contains all the top level nodes, and the unary relation
$L$ contains all the (former) leaves. It
is easy to write a $\Pi_t$-formula $\psi'(X)$ such that $\phi'$ has a
satisfying assignment of weight $k$ if, and only if, there exists a
$k$-element subset $B\subseteq C$ such that $\mathcal C\models\psi'(B)$.

\begin{figure}[ht]
\begin{center}
\includegraphics[width=5cm]{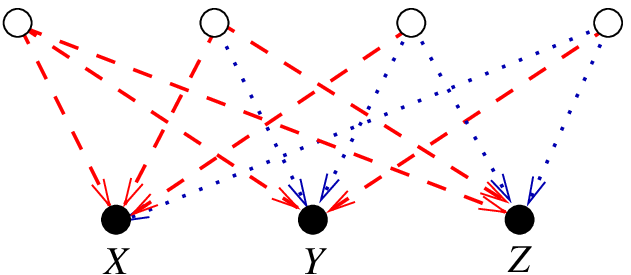}
\end{center}
\caption{}\label{fig:fag}
\end{figure}

This can best be illustrated with a simple example: Let 
\[
\phi':=(X\vee Y\vee Z)\wedge(X\vee\neg Y\vee Z)\wedge(X\vee \neg Y\vee\neg
Z)\wedge(\neg X\vee Y\vee\neg  Z).
\]
The corresponding structure $\mathcal C$ is displayed in Figure \ref{fig:fag}.

We let 
\begin{align*}
\psi'(X):=&\,\forall x(Xx\to Lx)\wedge\forall x\forall y_1\forall y_2\forall
y_3\Big(\big(Tx\wedge\bigwedge_{i=1}^3Exy_i\wedge\bigwedge_{1\le i<j\le
  3}y_i\neq y_j\big)\\
&\hspace{6cm}\longrightarrow\bigvee_{i=1}^3\big((Pxy_i\wedge
Xy_i)\vee(Nxy_i\wedge\neg Xy_i)\big)\Big).
\end{align*}

\medskip To prove $\mclass{\textup{FD}(\Pi_t)}\subseteq\W[t]$, we just note
that for every formula $\psi(X)\in\Pi_t$ and every structure $\mathcal A$,
there is a $C_{t,d}$-formula $\phi$ with the property that every assignment
$\alpha$ for $\phi$ corresponds to a set $B$, whose size is the weight
of the assignment, such that $\alpha$ satisfies $\phi$ if, and only if
$\mathcal A\models\psi(B)$. Here $d$ is a constant that just depends on
$\psi$. Furthermore, the transformation $(\mathcal A,\psi)\mapsto \phi$ is
computable in time polynomial in $\mathcal A$.  \proofend

We do not know of any simple proof of the equivalence between the two
characterizations of the W-hierarchy in terms of slicewise
$\Sigma_{t,u}$-definability (cf.\ Theorem \ref{theo:15}) and $\Pi_t$-Fagin definability. While the proof of
the previous theorem shows that $\Pi_t$-Fagin definability is actually quite
close to the definition of $\W[t]$ in terms of the weighted satisfiability
problem for $C_t$-formulas, it seems that it is a significant step to get from
there to slicewise $\Sigma_{t,u}$-definability.

Our last result is another characterization of the W-hierarchy in terms of
Fagin definability that is much closer to the slicewise characterization.
A first-order formula $\phi(X)$ is
\emph{bounded to} the $r$-ary relation variable $X$, if in
$\phi(X)$ quantifiers appear only
 in  the form 
$\exists x_{1} \ldots \exists x_{r}(Xx_{1} \ldots  x_{r} \wedge \psi)$ or
 $\forall x_{1} \ldots \forall x_{r}(Xx_{1} \ldots  x_{r} \rightarrow \psi)$,
 which we abbreviate by 
 $\exists \bar{x}\in X\;\psi$ and
 $\forall \bar{x}\in X\;\psi$, respectively.
 For  $t\ge 1$ we let $\Pi^b_t$ be the class of all 
formulas $\varphi(X)$ 
of the form
\[
\forall\bar x_1\exists\bar x_2\ldots Q\bar x_t\theta 
\]
where $Q=\forall$ if $t$ is odd and $Q=\exists$ otherwise and where $\theta$ 
is
bounded to  $X$.

For example, \emph{CLIQUE} is Fagin-defined by the $\Pi^b_0$-formula
$$
 \forall x\in X \forall y\in X(x\not=y \rightarrow Exy)
$$
and  \emph{DOMINATING SET} by the $\Pi^b_1$-formula
$$
 \forall x \exists y\in X(x=y \vee Exy).
$$

\begin{Theorem}\label{theo:17} 
For $t\ge 1$, $\W[t]=
\mclass{\textup{FD}(\Pi^b_{t-1})}$.
\end{Theorem}
\emph{Proof: } First, assume that the problem 
$P\subseteq\STR[\tau]\times\mathbb N$ is
Fagin-defined by $\phi(X)\in \Pi_{t-1}^b$, say
\[
\phi(X):=\forall \bar{y}_1 \exists \bar{y}_2 \forall \bar{y}_3 \ldots 
Q\bar{y}_t \psi,
\]
where $\psi$ only contains bounded quantifiers. Let $l$ be the maximum
of the lengths of the tuples $\bar y_i$, for $1\le i\le t$.  
For simplicity,  let  us assume that $X$ is unary.
Since $X y$ is equivalent to $\exists   z\in X \, z = y$, 
we can assume that in
$\psi$, the variable $X$ only occurs in quantifier bounds.
We show that
$P\in\W[t]$.  Given a parameter $k$ set (with new
variables $x_1,\ldots,x_k)$
\[
\phi^k:=\exists x_1\ldots \exists x_k\forall \bar{y}_1 \exists \bar{y}_2 
\forall \bar{y}_3
\ldots Q\bar{y}_t(\bigwedge_{1\le i< j\le k}x_i\neq x_j\wedge \psi^*),
\]
where $\psi^*$ is obtained from $\psi$ by inductively replacing
$\forall u\!\in \! X\chi(u)$ and $\exists u\! \in \!  X\,  \chi(u)$ by
$\bigwedge_{i=1}^k \chi(x_i)$ and $\bigvee_{i=1}^k \chi(x_i)$,
respectively. Note that $\phi^k$ is a $\Sigma_{t,l}$-formula and that
for every structure ${\mathcal A}$,
\[
({\mathcal A},k)\in P\iff{\mathcal A}\models
\phi^k.
\]
Thus, $P$ is slicewise $\Sigma_{t,l}$-definable and hence in $\W[t]$.

\medskip
For the converse direction, we prove that for all $t,u\ge 1$ the problem
$\MC(\Sigma_{t,u})|_{\GRA}$ is in $\textup{FD}(\Pi^b_{t-1})$.
The idea of this reduction is to associate with every graph $\mathcal G$ and
$\Sigma_{t,u}$-formula $\exists x_1\ldots\exists x_k\phi$ a structure $\mathcal C$ which essentially is a
Boolean circuit whose satisfying assignments of size $k$ correspond to
assignments to the variables $x_1,\ldots,x_k$ such that $\mathcal G$ satisfies
$\phi$. We can Fagin-define the weighted satisfiability problem for this
circuit by a $\Pi^b_{t-1}$-formula. We leave the details to the reader.

For readers familiar with \cite{dowfel99} (Theorem 12.6 on page 299 is the
relevant result), we state another proof. For $t=1$,
the result follows from the fact the $\W[1]$-complete problem \CLI\ is
$\Pi^b_0$-Fagin definable. For odd $t\ge 2$, the problem \textit{WEIGHTED
  ANTIMONOTONE $t$-NORMALIZED SATISFIABILITY} is
$\W[t]$-complete. It is parameterized m-reducible to the problem Fagin-defined
by the $\Pi_{t-1}^b$-formula
\begin{align*}
\phi (X):=
&\forall y_0 \forall y_1\exists y_2 \forall y_3 \ldots \exists y_{t-1}\\
&((Ey_{0}y_1\wedge Ey_1y_2\wedge\ldots \wedge Ey_{t-2}y_{t-1}) \rightarrow 
\forall x\in X\neg Ey_{t-1}x).
\end{align*}
For even $t$ we use the completeness of \textit{WEIGHTED
  MONOTONE $t$-NORMALIZED SATISFIABILITY} and argue similarly.
\proofend

\begin{Remark}
  Downey, Fellows and Taylor \cite{dowfeltay96} proved that the parameterized
  model-checking problem for full first-order logic is complete for the class
  $\text{AW}[*]$, a parameterized complexity class above the W-hierarchy that
  is defined in terms of the satisfiability problem for quantified Boolean
  formulas.
\end{Remark}


\end{document}